\documentclass[twocolumn]{aa}

\usepackage{amsmath}
\usepackage{amssymb}
\usepackage{acronym}
\usepackage{color}
\usepackage{graphicx}
\usepackage{mathrsfs}
\usepackage{multirow}
\usepackage{comment}
\usepackage[normalem]{ulem} \usepackage{soul}         

\usepackage{tensind}
\tensordelimiter{?}
\tensorformat{lrc}
\usepackage{pifont} %

\usepackage{natbib}
\usepackage[varg]{txfonts}

\usepackage{orcidlink}

\usepackage{hyperref}

\providecommand{\eprint}[1]{\href{https://arxiv.org/abs/#1}{#1}}

\newcommand{\ie}{\textit{i.e.}}
\newcommand{\eg}{\textit{e.g.}}
\newcommand{\be}{\begin{equation}}
	\newcommand{\ee}{\end{equation}}
\newcommand{\bea}{\begin{eqnarray}}
	\newcommand{\eea}{\end{eqnarray}}
\newcommand{\bel}{\begin{align}}
	\newcommand{\eel}{\end{align}}

\def\i{{\rm i}}

\def\GMc2{{\rm G M_{\odot} c^{-2}}}

\def\eps{\epsilon}

\def\knec{{\tt kNEC}}
\def\knecnn{{\tt kNECnn}}
\def\snec{{\tt SNEC}}
\def\skynet{{\tt SkyNet}}

\def\ENDF{\href{https://www-nds.iaea.org/exfor/endf.htm}{ENDF/B-VIII.0}}
\def\NIST{\href{https://www.nist.gov/pml/xcom-photon-cross-sections-database}{NIST-XCOM}}

\def\BLh{{\tt BLh}}
\def\DD2{{\tt DD2}}

\definecolor{cyan}{rgb}{0,0.9,0.9}
\definecolor{orange}{rgb}{0.9,0.5,0}
\definecolor{magenta}{rgb}{1,0,1}
\definecolor{purple}{rgb}{0.8,0.4,0.8}
\definecolor{gray}{rgb}{0.8242,0.8242,0.8242}

\newcommand{\newtxt}[1]{#1}

\begin{document}

\title{Impact of in situ nuclear networks and atomic opacities on neutron star merger ejecta dynamics, nucleosynthesis, and kilonovae}
\titlerunning{Impact of in situ NN and atomic opacities on BNSM ejecta dynamics, nucleosynthesis, and KN}

\author{Fabio Magistrelli\inst{1}\,\orcidlink{0009-0005-0976-7851} \and Sebastiano Bernuzzi\inst{1}\,\orcidlink{0000-0002-2334-0935} \and Albino Perego\inst{2,3}\,\orcidlink{0000-0002-0936-8237} \and Maximilian Jacobi\inst{1}\,\orcidlink{0000-0001-8168-4579} \and Christopher J. Fontes\inst{4,5}\orcidlink{0000-0003-1087-2964}}

\institute{
	Theoretisch-Physikalisches Institut, Friedrich-Schiller-Universit{\"a}t Jena, 07743, Jena, Germany \\
	\email{fabio.magistrelli@uni-jena.de} 	\and
	Dipartimento di Fisica, Università di Trento, Via Sommarive 14, 38123 Trento, Italy
	\and
	INFN-TIFPA, Trento Institute for Fundamental Physics and Applications, via Sommarive 14, I-38123 Trento, Italy 		\and
	Center for Theoretical Astrophysics, Los Alamos National Laboratory, Los Alamos, NM 87545, USA
	\and
	Computational Physics Division, Los Alamos National Laboratory, Los Alamos, NM, 87545, USA	}

\date{\today}

\abstract
    {Binary neutron star merger (BNSM) ejecta are key sites of rapid neutron capture ($r$-process) nucleosynthesis and produce kilonovae powered by the radioactive decay of freshly synthesized nuclei.
    Modeling their evolution requires multi-physics simulations involving hydrodynamics, nuclear reactions, and radiative processes.
    The impact of nuclear burning and atomic opacity is poorly understood and often treated with simplified prescriptions.} 
    {We systematically investigate different treatments of nuclear heating, particle thermalization, and atomic opacities in radiation-hydrodynamics simulations of BNSM ejecta and kilonova light curves.}
   {Ejecta profiles from long-term numerical-relativity simulations of asymmetric neutron star binaries with a massive neutron star remnant are evolved to $\sim 30$~days using a 2D ray-by-ray approach.
   	We compare simplified heating-rate and thermalization prescriptions with in-situ Nuclear reaction Network (NN) calculations that track nuclear energy deposition and include a composition-dependent thermalization scheme.
    We also contrast various gray opacity models with a frequency-dependent treatment based on atomic calculations.}
   {Coupling NN and hydrodynamics significantly affects nucleosynthesis and kilonova emission.    	Assuming homologous expansion alters abundance evolution and produces a narrower, less populated    	second $r$-process peak and a third peak shifted to higher mass numbers.     The back-reaction of nuclear heating affects the temperature evolution enough to delay and redden the early ($t\sim$~hours) kilonova peaks.     %
    A constant thermalization efficiency underestimates and reddens the early emission     while overestimating the late-time luminosity compared to the composition-dependent treatment.
    Analytical opacity prescriptions yield a more extended, colder photosphere, resulting in dimmer, redder kilonovae at early times ($t\lesssim$~hour), while the delayed recession of the photosphere prolongs the red emission at $t\gtrsim5$~days.
   }   
   {Coupling hydrodynamics to an in-situ NN is crucial for reliable nucleosynthesis and kilonova predictions.
   	Resolving the first several hundred milliseconds of the hydrodynamics is essential for robust nucleosynthesis calculations.
    Composition-dependent thermalization and frequency-dependent, atomic-physics-based opacities are needed to accurately capture the temperature evolution of the ejecta and the brightness and color evolution of the kilonova.
    Calibrated analytic nuclear-power fits with simplified thermalization and opacity prescriptions can still reproduce the density and temperature evolution of the ejecta.     
   }
\keywords{Stars: neutron --
	Nuclear reactions, nucleosynthesis, abundances --
	Transients: kilonovae --
	Radiation mechanisms: thermal --
	Opacity --
	Atomic data --
	Methods: numerical
}

\maketitle

\section{Introduction}
\label{sec:introduction}

Binary neutron star mergers (BNSMs) are one of the main sources for the production of heavy elements via the rapid neutron capture ($r$-process) nucleosynthesis \citep{Eichler:1989ve, Li:1998bw, Metzger:2010sy, Cowan:2019pkx, Perego:2021dpw}. The first multi-messenger observation of a BSNM was triggered by the gravitational wave (GW) signal GW170817, and associated with the short gamma-ray burst (GRB) GRB 170817A and the kilonova AT2017gfo, i.e. the electromagnetic (EM) counterpart powered by the decay of the freshly synthesized, unstable nuclei \citep{LIGOScientific:2017vwq, LIGOScientific:2017zic, GBM:2017lvd, Coulter:2017wya, Drout:2017ijr, Savchenko:2017ffs}.
Kilonova candidates have also been identified in association with other short GRBs \citep{Berger:2013wna, Gompertz:2017mbv, Rossi:2019fnm} and hybrid or long GRBs \citep{Rastinejad:2022zbg, Troja:2022yya}.

The properties of the matter ejected from a BNSM (mass, temperature, velocity, and electron fraction) are determined by the merger and post-merger dynamics, neutrino interactions, and the nuclear equation of state (EOS) \citep{Rosswog:2013kqa, Radice:2018pdn, Perego:2019adq, Foucart:2020qjb, Bernuzzi:2020tgt, Fujibayashi:2020dvr, Just:2021vzy, Kiuchi:2022nin, Lund:2024fjk, Cheong:2024cnb}.
The nuclear reactions burning in the expanding material depend on these initial conditions and on the thermodynamic evolution of the outflow.
The latter is dominated by its hydrodynamic expansion and affected by the released nuclear energy and radiative transport effects \citep{Korobkin:2012uy, Lippuner:2015gwa, Cowan:2019pkx, Fryer:2023osz, Magistrelli:2024zmk}.
The final EM (optical, UV, near-infrared, and $\gamma$) emission is dictated by the thermal energy of the ejecta, the nuclear power released in the reacting material, the way the products of the nuclear reactions thermalize with the surrounding matter, the ejecta morphology, and the atomic opacities in the outflow \citep{Barnes:2016umi, Kasen:2017sxr, Wollaeger:2017ahm, Kasen:2018drm, Hotokezaka:2019uwo, Tanaka:2019iqp, Korobkin:2019uxw, Fontes:2019tlk, Korobkin:2020spe, Gillanders:2022opm, Fontes:2022pue, Collins:2022ocl, Shingles:2023kua, Pognan:2024ise, Sneppen:2024jch}.
To understand the nucleosynthesis in BNSM ejecta, how they can participate in the galactic enrichment, and the associated EM counterparts, it is thus necessary to model all of these complex aspects of the merger and post-merger dynamics.

General-relativistic (magneto)-hydrodynamics (GRMHD) simulations implementing candidate nuclear EOSs can follow a binary system from the inspiral phase until hundreds of milliseconds after merger, collecting information on the thermodynamic properties of the outflows
\citep[\eg][]{Radice:2016dwd, Bauswein:2020aag, Nedora:2020hxc, Combi:2022nhg, Kiuchi:2023obe, Fields:2024pob}.
New moment-based neutrino-transport schemes and Monte Carlo methods have improved our understanding of the role of neutrinos in driving mass ejection (in particular for long-lived remnants) and setting the electron fraction, and thus initial composition, of these outflows
\citep{Foucart:2015gaa, Radice:2021jtw, Zappa:2022rpd, Kawaguchi:2024naa}.
Usually, ejecta properties extracted from numerical relativity (NR) simulations are transferred to other codes for their long-term (approximately days) evolution with the use of tracers or some suitable mapping \citep{Korobkin:2012uy, Martin:2015hxa, Barnes:2016umi, Radice:2018pdn, Perego:2020evn, Curtis:2023zfo, Groenewegen:2025ezj}.

As the material expands and cools, the temperature eventually drops below the nuclear statical equilibrium (NSE) threshold.
Beyond this point, an additional source of uncertainty in modeling $r$-process nucleosynthesis arises from the poorly constrained nuclear properties of isotopes far from stability.
In particular, simulating ejecta from BNSMs requires knowledge of the properties of nuclei near the neutron drip line.
Despite significant theoretical and experimental effort, key quantities such as nuclear masses, $\beta$-decay rates, neutron-capture cross sections, and fission yields remain highly uncertain, yet they strongly impact nucleosynthesis pathways and influence EM observables \citep{Mumpower:2015ova, Barnes:2020nfi, Zhu:2022qhc, Martinez-Pinedo:2023bkp, Mumpower:2024mlh, Martinet:2025nzx}.

Most of the current ejecta models treat nuclear physics and radiation-hydrodynamic separately.
For example, \cite{Rosswog:2013kqa, Kawaguchi:2020vbf, Wu:2021ibi, Kawaguchi:2024hdk} input precomputed, analytical nuclear powers fits in the energy equation. In \cite{Korobkin:2012uy, Radice:2018pdn, Perego:2020evn, Gillanders:2022opm, Just:2021vzy, Curtis:2023zfo}, the nucleosynthesis is calculated via a post-processing method by prescribing a simplified homologous expansion and feeding it to a Nuclear reaction Network (NN).
Similar simplifications have been used to perform full Monte Carlo radiative-transport simulations to calculate kilonova light curves and spectra \citep{Collins:2022ocl, Shingles:2023kua, Pognan:2024ise}.
These methods neglect the complex interplay between fluid dynamics, nucleosynthesis, and radiative transport, and they fail to capture the nonlinear feedback of radioactive heating and composition-dependent radiation-matter interactions on the outflow dynamics.
To address this limitation and study the interplay between ejecta dynamics and nucleosynthesis, \cite{Magistrelli:2024zmk} performed the first analysis of BNSM ejecta evolution that couples ray-by-ray radiation-hydrodynamic expansion with an in-situ NN.
This more self-consistent modeling revealed qualitative differences in the predicted kilonova emission and nucleosynthesis.
The same framework was applied in \cite{Bernuzzi:2024mfx} and \cite{Jacobi:2025eak} to investigate outflows from asymmetric BNSMs.

In this work, we assess the impact of in-situ nuclear-burning models and atomic opacities in radiation-hydrodynamics BNSM ejecta simulations. In order to explore efficiently different models, we focus on a 2D ray-by-ray setup where radiation hydrodynamics is treated in the diffusion limit.
In Sec.~\ref{sec:method}, we describe our frequency-dependent radiation-hydrodynamics setup and its coupling with the in-situ NN, which updates the \knecnn\ code~\citep[see also \cite{Morozova:2015bla,Wu:2021ibi}]{Magistrelli:2024zmk}.
The code implements simple, commonly used analytical prescriptions for the nuclear burning and opacities as well as different models for composition-informed thermalization and opacities enabled by the NN coupling.
We also discuss the extraction of initial ejecta profiles from NR simulations and describe how we incorporate the effects of a jet depositing energy into the polar regions of the ejecta.
Further details on the frequency-dependent radiative transport equations and on the initialization of the NN can be found in Appendices~\ref{app:rad_transp_eq}~and~\ref{app:res_NSEinit}.
In Sec.~\ref{sec:results}, we analyze the impact of each physical improvement introduced here relative to previous works.
We examine how the inclusion of an in-situ NN alters nucleosynthesis and light-curve predictions, and we asses the improvements enabled by the resulting composition-dependent thermalization and opacity models. We also study the impact of the additional energy deposited by a polar jet.
We summarize our findings in Sec.~\ref{sec:conclusion}.

\section{Method}
\label{sec:method}

In this work, we use \knecnn, a 2D ray-by-ray Lagrangian radiation-hydrodynamic code with in-situ NN, composition-dependent thermalization and opacities, and a frequency-dependent radiation transport scheme.
We based our development on the original \snec\ code from \cite{Morozova:2015bla} and its updated versions from \cite{Wu:2021ibi, Magistrelli:2024zmk}, extending the treatment of nuclear burning, thermalization, radiation transport, and opacities.

\subsection{Ray-by-ray radiation-hydrodynamics}
\label{subsec:rad-hydro}

The ejecta are evolved by solving the system of Lagrangian radiation-hydrodynamic equations presented in \snec\ \citep{Morozova:2015bla}.
Nonspherical effects in the dynamics are approximately incorporated under the axisymmetry and ray-by-ray assumptions.
The polar angle dependency of the ejecta properties is taken into account by mapping each angular section into an effective 1D, spherically symmetric problem.
The total mass of each section is scaled by the factor $\lambda_\theta = 4\pi / \Delta\Omega$, where $\Delta\Omega$ is the solid angle subtended by the angular section.
The sections are evolved independently, and the hydrodynamic variables and nucleosynthesis results are then recombined via mass weighted averages.
The mapping procedure assures that the intensive quantities are always preserved \citep{Magistrelli:2024zmk}. Note that non-radial flows of matter and radiation are neglected, as well as higher dimensional fluid instabilities (\eg\ Kelvin-Helmholtz), convection and the possibility of shells surpassing each other. These assumptions could in principle lead to overestimating the pressure work exchanged between the radial shells and introduce artificial shocks in the system.

In the 1D effective simulation of each angular section, the energy equation in spherical symmetry reads
\begin{equation}
	\label{eq:energy_cons}
	\frac{\partial \epsilon}{\partial t} = \frac{P}{\rho} \, \frac{\partial \ln\rho}{\partial t} - 4\pi r^2 Q \frac{\partial v}{\partial m} - \frac{\partial L}{\partial m} + \dot{\epsilon}_\text{nucl} \,,
\end{equation}
where $\epsilon$ is the specific internal energy, $t$ is time, and $P, \rho, r, v$ and $L$ are the pressure, density, radial position, radial velocity, and luminosity of the fluid element.
The Lagrangian coordinate is $m$, and $Q$ is the von Neumann-Richtmyer artificial viscosity \citep{VonNeumann:1950}.
The thermalized nuclear heating rate $\dot{\epsilon}_\text{nucl}$ accounts for the local production and partial thermalization of the energy released by all possible nuclear reactions occurring in the expanding material.
As already discussed in \cite{Magistrelli:2024zmk}, the EOS closing the set of hydrodynamic equations combines of the tabulated Helmholtz EOS \citep{Timmes:2000a, Lippuner:2017tyn} at high temperatures with the Paczynski EOS \citep{Paczynski:1986px, Morozova:2015bla, Wu:2021ibi} at lower temperatures, where the contribution of positrons is negligible.

The complete composition tracking implemented in \knecnn\ allows us to calculate the optical opacity on the fly accounting for the evolution of the density, temperature and composition profiles.
In our most sophisticated opacity model, presented in Sec.~\ref{subsec:opacity}, we make use of frequency-dependent opacities to calculate the luminosity in Eq.~\eqref{eq:energy_cons}, the position of the photospheres, and the observed kilonova fluxes.
The luminosity appearing on the right-hand side of Eq.~\ref{eq:energy_cons} is obtained as
\begin{equation}
	\label{eq:lum_eq}
	L = 4\pi r^2 \int_{0}^{\infty} F_\nu\, d\nu \,,
		\end{equation}
where the radiative flux for each shell is calculated under the assumption of local thermodynamical equilibrium (LTE) \newtxt{and in the single-temperature diffusion approximation} as
\begin{equation}
	\label{eq:flux_eq}
	F_\nu = -(4\pi r)^2 \frac{\lambda_\nu}{3 \bar{\kappa}_\nu} \frac{\partial B_\nu}{\partial T} \frac{\partial T}{\partial m} \,.
\end{equation}
Here, $B_\nu$ is the Planck spectrum, $\bar{\kappa}_\nu$ the composition-averaged opacity (see Sec~\ref{subsec:opacity} for further details), and $\lambda_\nu$ the flux-limiter
\begin{equation}
	\label{eq:flux_limiter}
	\lambda_\nu = \frac{6 + 3R_\nu}{6 + 3R_\nu + R_\nu^2} \,,
\end{equation}
with
\begin{equation}
\label{eq:R_factor}
	R_\nu = \frac{4\pi r^2}{\bar{\kappa}_\nu} \frac{\partial \log B_\nu}{\partial T} \left| \frac{\partial T}{\partial m} \right| \,,
\end{equation}
which is proportional to the one defined in \cite{Levermore:1981} (see Appendix~\ref{app:rad_transp_eq} for further details).

The numerical solution of Eq.~\ref{eq:energy_cons} requires an implicit solve involving the EOS and the temperature.
This is performed using a Newton-Raphson iteration on the temperature, nested inside a fixed-point iteration for the internal energy.
The frequency-dependent generalization follows the same algorithm, but each Newton–Raphson step additionally requires evaluating temperature derivatives of the blackbody function as well as computing integrals over all frequency groups.

To estimate the kilonova light curves, we first calculate the position $r_{\nu_\text{ph}}$ of each frequency-dependent photosphere by imposing
\begin{equation}
	\int_{r_{\nu_\text{ph}}}^\infty \tau_\nu(r) \, dr = \frac{2}{3} \,,
\end{equation}
where $\tau_\nu = \bar{\kappa}_\nu \rho$ is the optical depth. For thermal emission, the flux emitted by a spherically symmetric source is given by $f_\nu^S = \pi B_\nu$, where $B_\nu$ is the blackbody function.
The contribution from the photosphere to the flux measured by an observer at a distance $D$ is given by $f_\nu = f_\nu^S r_{\nu_\text{ph}}^2/D^2$.
Using the Stefan-Boltzmann law and the same arguments as in \cite{Wu:2021ibi} for the nuclear contributions from regions outside the photosphere, the observed kilonova fluxes can be expressed as
\begin{equation}
	\label{eq:fluxes_kn}
	f_\nu = \frac{1}{4 \pi D^2} ~
		\left( 4 \pi r_{\nu_\text{ph}}^2 F_{\nu,\nu_\text{ph}=\nu} + \int_{r_{\nu_\text{ph}}}^{r_\text{max}} \frac{\pi \dot{\eps}_{\text{nucl},m}}{\sigma T_m^4} B_\nu(T_m) \, dm \right) \,,
\end{equation}
where $F_{\nu,\nu_\text{ph}}$ is the flux at frequency $\nu$ at the photosphere for photons of frequency $\nu_\text{ph}$, $r_\text{max}$ is the outer boundary of the outflow, $\dot{\eps}_{\text{nucl},m}$ is the local nuclear heating rate, and $\sigma$ is the Stefan-Boltzmann constant.
The observed AB magnitudes are then calculated as in \cite{Wu:2021ibi} and recombined accounting for the viewing angle as in \cite{Martin:2015hxa, Perego:2017wtu}.

\subsection{Nuclear network coupling}
\label{subsec:nnCoupling}

We evolve the matter composition and compute the specific energy deposition $\dot{\epsilon}_\text{nucl}$ self-consistently, by providing each fluid element in our simulation with an in-situ NN.
In the current version of the code, we use an implementation of the \skynet\ NN \citep{Lippuner:2017tyn} which we adapted to interact with the kilonova-version of \snec\ from \cite{Wu:2021ibi}.
The coupling infrastructure is general and in principle able to host other NNs. The NN used in this paper includes 7836 isotopes up to~$^{337}$Cn and uses the JINA REACLIB \citep{Cyburt:2010a} database and the same setup as in \cite{Lippuner:2015gwa, Perego:2020evn, Magistrelli:2024zmk}.

Within every hydrodynamic time-step $\Delta t$, each fluid element independently evolves the associated NN, which automatically defines its own sub-steps based on the nuclear timescales.
We call $t_0$ the initial time of the hydrodynamical time-step.
During the NN evolution, the density is prescribed by a log-interpolation between $\rho(t_0)$ and $\rho(t_0+\Delta t)$. The latter is calculated from $\rho(t_0)$ and the velocity of the shell boundaries updated via the momentum transport equation (for which no nuclear input is required).
During the timestep, we assume a constant $T = T(t_0)$ for the NN calculations (no self-heating).
The temperature will then be updated by the second part of the hydrodynamic step.
We collect the energy released by nuclear reactions during each sub-step, and define the (non-thermalized) nuclear power $\dot{q}_\text{nucl}(t_0)$ as the total energy produced, divided by $\Delta t$. This quantity will enter the energy equation~\eqref{eq:energy_cons} after being thermalized into $\dot{\epsilon}_\text{nucl}(t_0)$ as described in Sec.~\ref{subsec:thermalization}.
At the end of the NN calculations, we use the updated isotopic composition to get the mean molecular weight, which enters the EOS, and the opacity profile, needed to calculate the luminosity term in Eq.~\eqref{eq:energy_cons} and the observed fluxes. In other words, we use the abundances at $t=t_0 + \Delta t$ as a representation of the composition during the whole hydrodynamic step.
This procedure produces all the information needed to evolve Eq.~\eqref{eq:energy_cons}, and thus to update temperature and \newtxt{specific internal energy} of the outflow.

Our ab-initio NR simulations do not track nor evolve the detailed matter composition, but always rely on EOS tabulated for matter in nuclear statistical equilibrium (NSE). We must thus prescribe the isotopic composition at the beginning of the radiation-hydrodynamical simulation.
At merger, the colliding matter from the two neutron stars typically reaches a high enough temperature ($T\gtrsim8$~GK) to ensure hot NSE conditions.
However, expansion usually causes the unbound material to drop out of NSE before reaching the extraction radius.
In particular, many fluid elements recorded at the extraction radius have $T < 5$~GK in the initial ejecta profile (see Fig.~\ref{fig:initial_profiles}).
If tracers are modeled inside the NR simulation, or if they can be reconstructed in post-processing, one can study the evolution of these fluid elements starting from their reconstructed thermodynamics trajectories. In this case, one can post-process a NN on these trajectories and predict the composition of the fluid elements at the extraction radius properly accounting for the out-of-NSE transition. The operation is not self-consistent, as it does not include out-of-NSE effects in the original dynamic simulation. However, it gives a reliable estimate for the composition of the ejecta at the extraction radius that can be used to initialize \knecnn.
If the necessary information to define tracers is missing, our code relies on the NSE assumption to estimate the initial isotopic composition.
This is the case for the NR simulations considered in this work.

The simplest initialization method implemented in \knecnn, and already used in \cite{Magistrelli:2024zmk}, assumes NSE at any temperature $T_0$ registered at the extraction radius.
We will refer to this procedure as cold-NSE initialization (\texttt{CNSE}) in the rest of this work.
For each tracked isotope, the initial abundance is given by
\begin{equation}
	\label{eq:NSEcomp}
	Y_i = \frac{n_i}{n_b} \propto Y_p^{Z_i} \, Y_n^{(A_i - Z_i)} \, \frac{\rho^{A_i - 1} \, e^{\text{BE}_i/k_\text{B}T_0}} {T_0^{\,3(A_i-1)/2}} \,,
\end{equation}
where $n_b$ is the baryon number density, $n_i$, $A_i$, $Z_i$, and $\text{BE}_i$ are the number density, mass number, atomic number and binding energy of the $i$-th isotope, $Y_p$ and $Y_n$ are the abundances of free protons and neutrons, and $k_\text{B}$ is the Boltzmann constant \citep{Cowan:2019pkx, Perego:2021dpw}.
The initial isotopic mass fraction is given by $X_i = A_i Y_i$.
The mass and charge conservation, $\sum_i A_i Y_i = 1$ and $\sum_i Z_i Y_i = Y_e$, with $Y_e$ electron fraction, constrain the NSE composition, which thus depends on $\rho$, $T$ and $Y_e$.
See \cite{Lippuner:2017tyn} for details about \skynet's implementation of the NSE condition.

For matter that experienced hot NSE during the NR simulation, but underwent NSE freeze-out before reaching the extraction radius, the Boltzmann term in Eq.~\eqref{eq:NSEcomp} tends to overestimate the abundances of the iron-group elements.
Moreover, the impact of nuclear reactions happening immediately after NSE drop-out is neglected.
To correct these effects, we implement a backtracking-NSE initialization similar to the one described in \cite{Radice:2016dwd, Perego:2020evn}.
Essentially, if $T_0$, $\rho_0$, $Y_{e,0}$, $s_0$ are the properties at the extraction radius, we assume that the fluid element had expanded adiabatically from its NSE configuration at some $T_\text{NSE} > T_0$, $\rho_\text{NSE}$, $Y_{e,\text{NSE}}$, $s_\text{NSE}$.
On the small timescales ($\lesssim 10$~ms) needed for the ejected matter to get to the extraction sphere, we expect the electron fraction and the entropy not to change significantly, so $Y_{e,\text{NSE}} \simeq Y_{e,0}$, $s_\text{NSE} \simeq s_0$.
We check a posteriori that the relative changes in the entropy and electron fraction are on the order of 5\% or less during the pre-evolution phase in most part of the ejecta (see Fig.~\ref{fig:bt_hist}).
We finally estimate the density $\rho_\text{NSE}$ of the fluid element at NSE freeze-out by using the EOS employed in the NR simulation and starting from $T_\text{NSE}$, $Y_{e,\text{NSE}} = Y_{e,0}$, $s_\text{NSE} = s_0$.
We calculate the correspondent NSE composition via Eq.~\eqref{eq:NSEcomp}, and then mimic the abundance evolution up to the extraction radius assuming homologous expansion. In particular, we parametrize the density evolution as in Eq.~(1) of \cite{Lippuner:2015gwa}, and estimate the expansion timescale $\tau_\text{ex}$ as in \cite{Radice:2016dwd}\footnote{
	The prescription of \cite{Lippuner:2015gwa} is meant to approximate the physical evolution of a fluid element starting at early times after the ejection. The assumption of homologous expansion, not realistic at early times, is needed to get an estimate of the expansion timescale of the ejecta. Since variations of a factor of a few in $\tau_\text{ex}$ do not significantly affect the evolution of the NN \citep{Lippuner:2015gwa, Perego:2021dpw}, our assumption gives a reliable approximation for the composition at the extraction radius.}.
To ensure that the evolved thermodynamic trajectory matches the physical conditions of the fluid element at the extraction radius, we also use an analytical prescription for the temperature history inspired by the adiabatic expansion law for an ideal gas, $T(\rho) = T_\text{NSE}\,(\rho(t) / \rho_\text{NSE})^{(\gamma -1)}$.
The effective adiabatic factor $\gamma$, specific for each fluid element, is determined by $\gamma = 1+ \ln(\rho_0 / \rho_\text{NSE}) / \ln(T_0 / T_\text{NSE})$, which ensures $T(\rho_0) = T_0$.
The composition at the end of the backtracking is used together with the originally extracted $s$ to define the initial condition of the fluid element.

Low-entropy tidal ejecta (more efficiently produced for strongly asymmetric binaries) are expelled before the merger can raise the temperature, and are therefore expected to retain a composition close to the one set by the cold-NSE condition, for which $T \lesssim 1$~MeV.
In principle, this would suggest using the cold-NSE method. However, in an NR simulation, numerical artifacts (\eg\ shock heating propagating inward from the surface of the compact objects) can boost the initial temperature of the neutron stars up to few MeV. In all the NR simulations analyzed in this paper, the neutron stars reach this temperature in $\sim1$~ms during the inspiral phase.
The spurious high temperatures introduce small artifacts in the tracers calculations and in the initialized NSE composition, regardless of the method used.
In Appendix~\ref{app:res_NSEinit}, we investigate the systematic uncertainties propagating from the different initialization methods to the nucleosynthesis and kilonova results.

\subsection{Nuclear power fits}
\label{subsec:nucl_pow_fits}

To investigate the impact of the in-situ NN, we need benchmark simulations using only ex-situ calculations.
For these simulations (labeled in what follows as \texttt{Apr2}) we calculate the nuclear power with an updated version of the analytical fits from \cite{Wu:2021ibi}.
At early times, $t \lesssim 0.1$~days, we consider the fitting function
\begin{equation}
	\dot{q}_\text{nucl} (t) = q_1 e^{-t/\beta} + q_0 \left[\frac{1}{2} - \frac{1}{\pi} \arctan\left(\frac{t - t_0}{\sigma}\right)\right]^\alpha \,,
\end{equation}
where $q_1$, $\beta$, $q_0$, $\alpha$, $t_0$ and $\sigma$ are fitting parameters.
The additional exponential term effectively generalizes the formula proposed by \cite{Korobkin:2012uy} to include initially mildly neutron-rich ($0.4 \lesssim Y_{e,0} < 0.5$) and proton-rich ($Y_{e,0} > 0.5$) fluid elements.
These trajectories, which do not produce $r$-process elements, typically exhibit an early fast drop in the associated nuclear power, followed by a return to the approximate $\arctan$ behavior (although usually shifted down by several orders of magnitude compared to an equivalent more neutron-rich fluid element).
At later times, $t \gtrsim 0.1~$days, we use the power law fit
\begin{equation}
	\dot{q}_\text{nucl} (t) = q_0' t^{-\alpha'} \,,
\end{equation}
where $q_0'$ and $\alpha'$ are additional fit parameters.
As in \cite{Wu:2021ibi}, the two nuclear powers are joined together by a log-scaled smoothing procedure applied between $10^3 \leq t \, [\textrm{s}] \leq 4 \times 10^4$ and centered on $t\sim0.1$~days.
The two fits combined describe the nuclear power over times after merger between $0.1$~s and $50$~days.

The original fits from \cite{Wu:2021ibi} relied on the nuclear power grid presented in \cite{Perego:2020evn}.
We perform our fits on the grid obtained in \cite{Chiesa:2023jno}, which improves on the original one presented in \cite{Perego:2020evn} by initializing the NN at a higher NSE temperature, $T_\textrm{NSE} = 8$~GK, to better capture NSE drop-out.
Additionally, specifically for this work, we have extended this grid by increasing the initial electron fraction, entropy, and expansion timescales ranges to $0.01 < Y_{e,0} < 0.6$, $1.4925 < s \, [k_B \, \textrm{baryon}^{-1}] < 300$, $0.5 < \tau \,[\textrm{ms}] < 200$, respectively.
Note that \cite{Wu:2021ibi} ceil the initial electron fraction to $Y_{e,0} = 0.48$ and use the table boundaries for higher values.
The NN is set up as described in Sec.~\ref{subsec:nnCoupling}, consistent with the rest of the calculations in this work.

\begin{figure}
	\centering
		\includegraphics[width=0.5\textwidth]{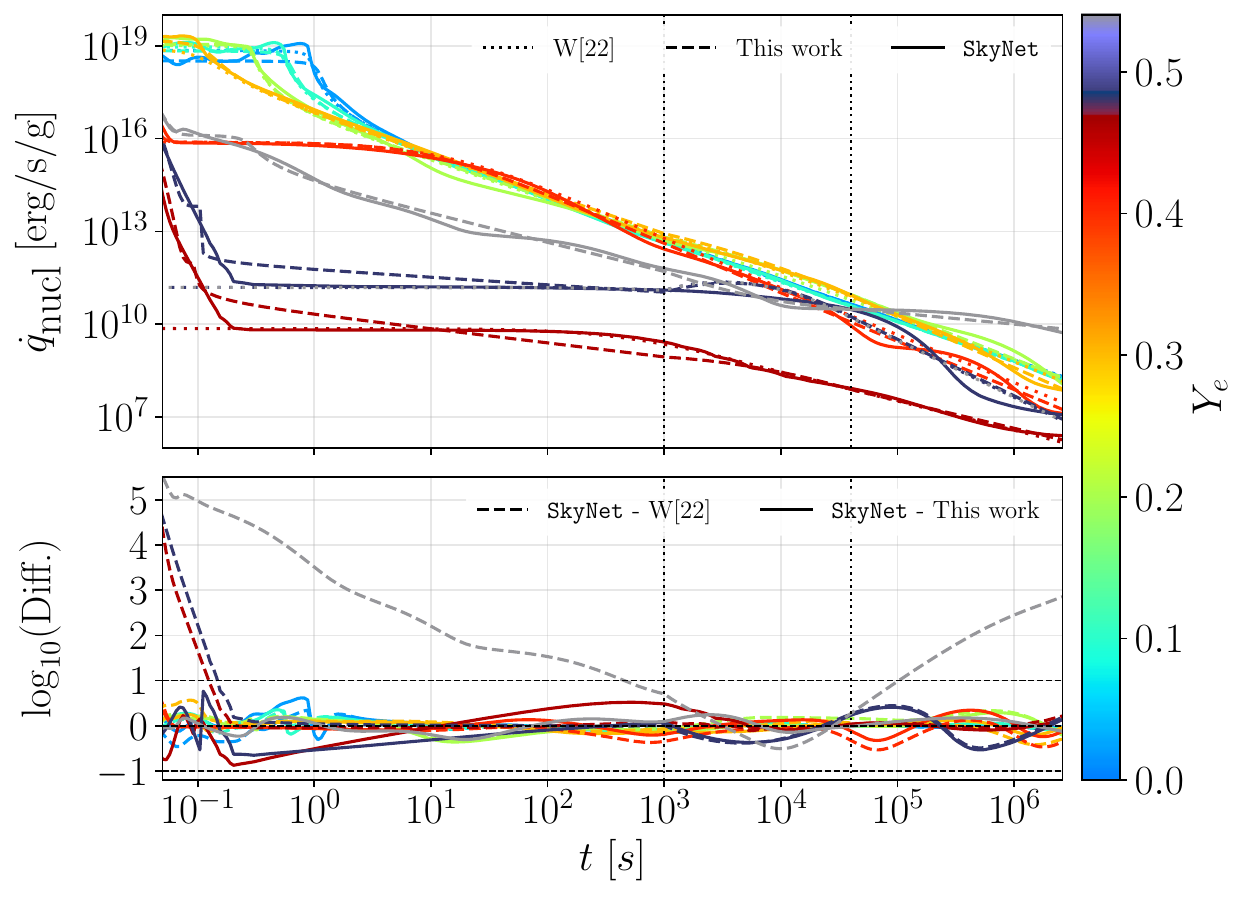}
	\caption{Nuclear power as a function of time. All the trajectories have initial entropy $s_0 \simeq 11~k_B~\textrm{baryon}^{-1}$, expansion timescale $\tau_0 \simeq 8.6$~ms, and electron fraction $Y_e \in [0.02, 0.1, 0.2, 0.3, 0.4, 0.46, 0.48, 0.54]$ indicated by the color map.
		The vertical dashed lines mark the transition between the early- and late-times fits.
		\textit{Top panel}: \skynet\ calculations (solid lines) and estimate from the updated (dashed lines) and old (dotted lines) nuclear power fits, respectively from this work and \cite{Wu:2021ibi}.
		\textit{Bottom panel}: Relative differences between the results from \skynet\ and from the new (solid lines) and previous (dashed lines) version of the fits.
		The dashed horizontal lines mark differences of one order of magnitude. 		}
	\label{fig:nucl_pow_fits}
\end{figure}

In Fig.~\ref{fig:nucl_pow_fits}, we compare the nuclear power calculated with \skynet\ against the predictions from the original fits of \cite{Wu:2021ibi} and our updated version.
The old fits fail to capture the very early-time energy production of the mildly neutron-rich and proton-rich matter, as well as the late-time nuclear power for proton-rich ejecta, which constitute most of the outflow for angles $30 \lesssim \theta~[\textrm{degrees}] \lesssim 60$, by several orders of magnitude.
The new fits only perform worse than the old ones in localized time intervals (\eg\ $1 < t~[\textrm{s}] < 10^3$ for $Y_e=0.46, 0.48$).
The discrepancy with the original \skynet\ calculations remains within a factor of order unity, assuring an overall improvement of the original fits of \cite{Wu:2021ibi}.
While a precise estimate of the nuclear power is not needed at early times, when the dynamics are dominated by the hydrodynamic explosion, it becomes crucial at late times to obtain reliable predictions of kilonova light curves.
In 3D ejecta simulations, where nuclear heating can affect the angular dynamics of the ejecta for timescales $t \gtrsim 1$~s, more accurate nuclear power fits might be necessary to capture the correct evolution of the outflow.

\subsection{Thermalization}
\label{subsec:thermalization}

In the simple thermalization scheme implemented in the kilonova version of \snec\ by \cite{Wu:2021ibi}, the thermalized heating rate is approximated by $\dot{\eps}_\text{nucl} = f_\text{th}^\text{\knec} \, \dot{q}_\text{nucl}$, where $\dot{q}_\text{nucl}$ is the total energy released by nuclear reactions, and $f_\text{th}^\text{\knec}$ is a constant and uniform thermalization coefficient.
The standard value $f_\text{th}^\text{\knec} = 0.5$ is adopted to represent the average behavior of all particles emitted by nuclear decays, ranging from easily escaping neutrinos to heavy, efficiently thermalizing, charged particles (fission fragments).
In what follows, we refer to this thermalization scheme as \texttt{ST05}.

In \knecnn, we implement time- and composition-dependent thermalization coefficients separately for electrons and positrons from $\beta^\pm$ decays, and for $\alpha$ particles and $\gamma$ rays emitted from all possible decays, including the $m_e \simeq 511~$keV photons from electron-positron annihilations.
We explicitly remove the energy carried away by (electron) neutrinos and antineutrinos emitted by $\beta^\pm$ decays from the thermalized heating rate.
We will denote this model as \texttt{DT08} in the rest of the paper.

We calculate the thermalization efficiency $f_\text{th}^{\alpha,\beta} (t)$ for $\alpha$ particles and electrons with the analytical prescriptions of \cite{Kasen:2018drm}. We assume that the same expression derived for electrons also holds for positrons.
The decay energy injected in these particles is calculated as $\dot{q}_k (t) = \sum_j \langle \dot{q}_j^k (t) \rangle$, where $\langle \dot{q}_j^k (t) \rangle$ is the average energy injected by the $j$-th decaying isotope in particles of the radiation type $k = \alpha, e^-, e^+$.
We calculate this quantity as
\begin{equation}
	\label{eq:rad_energy}
	\langle \dot{q}_j^k (t) \rangle = \frac{\langle E_j^k \rangle Y_j (t)}{m_p \, \tau_j} \,,
\end{equation}
where $m_p$ is the proton mass, $\tau_j$ is the average lifetime of the $j$-th isotope, and $\langle E_j^k \rangle$ is the mean energy released in particles of type $k$ after a radioactive decay.
The average lifetime accounts for all possible decays of the isotope, and the mean released energy averages over all possible decays accounting for their branching ratios.
Both these quantities are taken from the \ENDF\ database \citep{Brown:2018jhj}. We also include the free-neutron decay information from the IAEA nuclear data service website\footnote{
	\url{https://www-nds.iaea.org/relnsd/vcharthtml/VChartHTML.html}}.
The same expressions for $\langle \dot{q}_j^k (t) \rangle$ and Eq.~\eqref{eq:rad_energy} generalize for $k=\nu$, and prompt $\gamma$-rays.

\knecnn\ internally checks that the total rate of emitted energy $\langle \dot{q} \rangle = \sum_{k,j} \langle \dot{q}_j^k (t) \rangle$ does not exceed the nuclear power $\dot{q}_\text{nucl}$ calculated by the NN, and rescales all the $\dot{q}_k = \sum_j \langle \dot{q}_j^k (t) \rangle$ by the factor $\dot{q}_\text{nucl}/ \langle \dot{q} \rangle$ in case this happens.
We then calculate the power emitted in $\gamma$-rays from positron-electron pair annihilations (not included in the $\dot{q}_\text{nucl}$ from the NN), as $\langle \dot{q}_j^\text{pair} (t) \rangle = 2 m_e \langle n_{e^+} \rangle$, where $\langle n_{e^+} \rangle$ is the average number of emitted positrons reported in the database.

The thermalization factor $f_j^\gamma (t)$ for prompt $\gamma$-rays is calculated, as in \cite{Hotokezaka:2019uwo, Combi:2022nhg}, starting from the detailed composition of the ejecta and the same effective opacity tables from the \NIST\ catalog \citep{Berger:2010} used in \cite{Barnes:2016umi}.
The thermalization efficiency associated with the $\gamma$ particles emitted by nuclei of the $j$-th isotope through radioactive decays is given by
\begin{equation}
	\label{eq:therm_fact_gamma_j}
	f_j^\gamma(t) = 1 - \exp\left[ - \kappa_j^\gamma \Sigma_m (t) \right] \,,
\end{equation}
where $\kappa_j^\gamma$ is the absorptive opacity and $\Sigma_m (t)$ is the mass-weighted column density of the ejecta. In spherical symmetry,
\begin{equation}
	\Sigma_m(t) = \int_{m_0}^{m_\text{max}} \frac{dm}{M_\text{ej}} \int_m^{m_\text{max}} \frac{dm'}{4\pi r^2(m')} \,,
\end{equation}
where $M_\text{ej}$ and $m_0$ are the total mass of the ejecta and the mass of the remnant (which coincides with the Lagrangian coordinate of the first simulated shell) and $m_\text{max} = m_0 + M_\text{ej}$.
The absorptive opacity can be written as
\begin{equation}
	\label{eq:abs_opacity}
	\kappa_j^\gamma = \sum_k h_{k,j}^\gamma \, \kappa_{\text{eff}\,}^\gamma (E_{k,j}) \,,
\end{equation}
where $h_{k,j}^\gamma$ is the fraction of energy emitted in the $k$-th $\gamma$-ray, of energy $E_{k,j}$, by a decay of the $j$-th isotope \citep{Hotokezaka:2019uwo}, and $\kappa_{\text{eff}\,}^\gamma (E_{k,j})$ is the effective absorptive opacity. We read this information again from the \ENDF\ database \citep{Brown:2018jhj}, where the injection energy distributions are already averaged, for each isotope, over all accessible decays, accounting for their branching ratios.
The effective opacity is given by the mass-weighted average
\begin{equation}
	\label{eq:abs_opacity_eff}
	\kappa_\text{eff}^\gamma (E_{k,j}) = \sum_Z X_{Z} \, \kappa_{\text{eff},Z\,}^\gamma(E_{k,j}) \,,
\end{equation}
where $Z$ runs over all possible atomic numbers, $X_Z = \sum_{s:Z_s = Z} X_s$, with $s$ running through all the isotopes considered in the NN, and $\kappa_{\text{eff},Z}^\gamma (E_{k,j})$ is the effective opacity for a stopping material made only of isotopes with atomic number $Z$.
The latter is included in the \NIST\ catalog.
Since this database only contains information up to $Z=100$, we fix $ \kappa_{\text{eff},Z>100}^\gamma (E_{k,j}) = \kappa_0$ and use Eq.~\eqref{eq:abs_opacity_eff} to rewrite Eq.~\eqref{eq:abs_opacity} as
\begin{equation}
	\kappa_j^\gamma = \sum_{k,Z\le100} X_{Z} \, h_{k,j}^\gamma \, \kappa_{\text{eff},Z\,}^\gamma (E_{k,j}) + X_{\text{rem}\,} \kappa_0 \,,
\end{equation}
where $X_\text{rem} = 1-\sum_{Z\le100} X_Z$ is the cumulative mass fraction of the elements excluded from the database.
Assuming that the energy released in $\gamma$-rays is immediately thermalized, one gets
\begin{equation}
	\label{eq:therm_eps_gamma_nucl}
	\dot{\eps}_\text{nucl}^\gamma = \sum_j f_j^\gamma(t) \, \langle \dot{q}_j^\gamma (t) \rangle \,.
\end{equation}
where $\langle \dot{q}_j^\gamma (t) \rangle$ is given by Eq.~\eqref{eq:rad_energy} with $k=\gamma$.
For the simulations performed in this work, we fix $\kappa_0 = 0.1$~cm$^2$~g$^{-1}$.
Equations \eqref{eq:therm_fact_gamma_j}-\eqref{eq:therm_eps_gamma_nucl} also apply for $\gamma$-rays for the positron-electrons pair annihilations, for which $h_{k,j}^\text{pair} = \delta(E_{k,j} - m_e)$ and $\langle E_j^\text{pair} \rangle = 2m_e \text{BR}_{\beta^+}$, with $m_e$ electron mass and $\text{BR}_{\beta^+}$ branching ratio of the $\beta^+$ decay.

For the remaining nuclear power, associated with the recoil energy of daughter nuclei, fission fragments, (soft) $X$-rays, proton and neutron emissions, and delayed electron (\eg\ Auger) emission, we assume a constant and uniform thermalization factor $f_\text{th}^\text{oth}$.
The heavy charged particles are expected to thermalize efficiently at all times, leaving only to $X$-rays, injected nucleons and delayed electrons with a nontrivial behavior. In our simulations, we choose, as a standard value, $f_\text{th}^\text{oth} = 0.8 > f_\text{th}^\text{\knec}$, as we are now explicitly removing the neutrino contributions.
The total thermalized heating rate is finally given by
\begin{equation}
	\label{eq:nuclear_heating}
	\dot{\eps}_\text{nucl} = \dot{\epsilon}_\text{nucl}^\alpha + \dot{\epsilon}_\text{nucl}^{\beta} + \dot{\epsilon}_\text{nucl}^\gamma + \dot{\epsilon}_\text{nucl}^\text{pair} + \dot{\epsilon}_\text{nucl}^\text{oth} \,,
\end{equation}
where $\dot{\epsilon}_\text{nucl}^k = f_\text{th}^k \, \dot{q}_k$ for $k \in [ \alpha, \gamma, \textrm{pair}, \textrm{oth} ]$, $\dot{\epsilon}_\text{nucl}^\beta = f_\text{th}^\beta \, (\dot{q}_{e^-} + \dot{q}_{e^+})$, and $\dot{q}_\text{oth} = \dot{q}_\text{nucl} - \dot{q}_\alpha - \dot{q}_{e^-} - \dot{q}_{e^+} - \dot{q}_\gamma - \dot{q}_\nu$.
The $e^+-e^-$ pair annihilation photons are not included in the heating rate $\dot{q}_\text{nucl}$ from the NN, and thus their contribute must not be removed from $\dot{q}_\text{oth}$.

\subsection{Opacities}
\label{subsec:opacity}

In \knecnn, we implement several approaches to calculate the optical opacity.
The simplest model (named W[22]) uses gray opacities.
By plugging the standard definition of the Rosseland mean opacity into Eq.~\eqref{eq:lum_eq}, the radiative luminosity is
\begin{equation}
	\label{eq:lum_eq_ross}
	L_R = - (4\pi r^2)^2 \frac{\lambda a c}{3 \kappa_R} \frac{\partial T^4}{\partial m} \,,
\end{equation}
where $\lambda$ is an equivalent flux limiter as the one used in \cite{Morozova:2015bla, Wu:2021ibi} (see Appendix~\ref{app:rad_transp_eq} for further details). The simplest model is the same gray opacity prescription as in \cite{Wu:2021ibi}, where $\kappa_R = \kappa(Y_{e,0})$ has a constant-in-time profile depending only on the initial electron fraction.
The general physical idea behind this assumption is that, within each fluid element, a lower $Y_{e,0}$ will lead to the production of more $r$-process elements, resulting in a higher opacity.
This is a strong approximation, as it does not use information about the actual ejecta composition or its dynamics. 
This opacity model was employed in \cite{Magistrelli:2024zmk, Bernuzzi:2024mfx, Jacobi:2025eak}.

As a first, direct improvement of the analytical expression from \cite{Wu:2021ibi}, we implemented in \knecnn\ the opacity model of \cite{Just:2021vzy}.
In this model (referred to in what follows as J[22]) the opacity is a time-dependent analytical function of the cumulative lanthanide mass fractions and the gas temperature.
This models is based on the light curves calculated in \cite{Kasen:2017sxr}, which rely on radiative-transfer equations and atomic-physics calculations of the opacities of lanthanides and iron-group elements.
The prescription of \cite{Just:2021vzy} approximately reproduces the qualitative behavior of the kilonova from the original model.
It directly connects the production of lanthanides with a boost in the effective opacity, and it approximately includes the effect of electronic recombination at small temperatures ($T \lesssim 2000$~K).

For models based on the full tracking of the matter composition, we rely on the line-binned opacity calculations presented in \cite{Fontes:2019tlk, Fontes:2022pue}, and available online via the NIST-LANL Opacity Database \citep{Ralchenko:2025}.
In these publications, the population distribution of the ionization states is computed for a range of temperatures and densities using the Saha equation.
Detailed ab-initio atomic-structure calculations are then employed to compute the opacity as a function of discretized frequency intervals.
No assumptions are needed about the dynamical expansion.
\knecnn\ implements data for Cr, Se, Br, Zr, Pd, Te, Ce, Nd, Sm and U. 
In this work, we focus on the effect of using atomic data for lanthanides and actinides, and use only the resulting opacities for Nd and U to effectively define a composition-averaged opacity of the ejecta.
We discretize the frequency domain $\nu \in [3 \times 10^9, ~ 3.6 \times 10^{19}]$~Hz in $n_\textrm{gr}=300$ groups.
We use a piecewise logarithmic frequency grid with coarse spacing at low ($\nu \lesssim 2.42\times10^{12}$~Hz) and high ($\nu \gtrsim 7.25\times10^{15}$~Hz) energies, and higher resolution in the central energy region to better resolve spectral features.
We then calculate the group opacity from the line-binned results as prescribed in \citep{Fontes:2019tlk}.
The database includes data for $10^{-20} < \rho[\text{g}/\text{cm}^3] < 10^{-4}$ and $116 \lesssim T[\text{K}] \lesssim 5.8\times10^4$.
For densities or temperatures above the limits, we use the analytical model from \cite{Just:2021vzy}. Close to the upper boundaries of the table, we use an activation function to smoothly switch between the two models. For densities or temperatures below the limits, we use the opacity data at the lower edge of the table.
We use Nd and U data as representatives for all the lanthanides or actinides, respectively.
All the remaining elements are assigned an opacity $\kappa_\text{oth} = 0.2~k_T$~cm$^2$g$^{-1}$, with $k_T = 1$ if $T>2000$~K and $k_T = (T[\textrm{K}]/2000)^5$ for $T<2000$~K, consistent with the no-lanthanides case of the \cite{Just:2021vzy} approximation.
To get the composition-averaged opacity $\bar{\kappa}_\nu$ of Eq.~\eqref{eq:flux_eq}, we first estimate the opacity $\kappa_{\nu,i}(\rho, T)$ of each representative element by interpolating its tabulated values in the $\rho-T$ plane.
Then, for each frequency group, we average the contributions of the representative elements, weighting each element by the total mass fractions of the species it represents.
The composition-averaged opacity is given by
\begin{equation}
	\label{eq:kappa_represent}
	\bar{\kappa}_{\nu\,} (\rho, T\, |\, t) = \sum_i \sum_j X_{i,j}(t)\, \kappa_{\nu,i\,}(\rho, T\, |\, t) + X_\text{oth}(t) \, \kappa_\text{oth}(T) \,,
\end{equation}
where $X_\text{oth} = 1 - \sum_{i,j} X_{i,j}$.
In the sum, the index $i$ runs over the representative elements, while $j$ runs over all elements that are mapped to the $i$-th representative.

As a first gray model (\texttt{LANL-R}), we use again the Rosseland luminosity $L_R$ defined in Eq.~\eqref{eq:lum_eq_ross}, but we calculate $\kappa_R$ directly from the line-binned database as
\begin{equation}
	\label{eq:rosseland_op}
	\kappa_R = \left[ \frac{\sum_\nu \bar{\kappa}_\nu^{-1}\, \partial_T B_\nu\, \Delta\nu}{\sum_\nu \partial_T B_\nu\, \Delta\nu} \right]^{-1} \,.
\end{equation}
We also implement an analogous model (\texttt{LANL-P}) with the Planck gray opacity
\begin{equation}
	\kappa_P = \frac{\sum_\nu \bar{\kappa}_\nu B_\nu(T)\, \Delta\nu}{\sum_\nu B_\nu(T)\, \Delta\nu} \,.
\end{equation}
The Rosseland approximation, weighting more the transparent frequencies, is more suited for optically thick material. Its definition directly comes from the LTE and diffusion-approximation limits, leading to Eq.~\eqref{eq:flux_eq}.
The Planck gray opacity, weighting more the opaque frequencies, works better in an optically thin environment.
Both approximations are still not realistic, as gray opacities always assume thermal transport of radiation. In BNSM ejecta, the detailed structure of atomic energy levels and the corresponding absorption and emission lines play a crucial role in how radiation interacts with the material. Our most elaborate opacity model (\texttt{LANL}) explicitly includes the information from the opacity database into Eq.~\eqref{eq:lum_eq}, discretized as
\begin{equation}
	L = 4\pi r^2 \sum_{j_\textrm{gr}} F_{j_\textrm{gr}}\, \delta\nu_{j_\textrm{gr}} + L(\nu < \nu_1) + L(\nu > \nu_{n_\textrm{gr}}) \,,
\end{equation}
where $j_\textrm{gr} = 1,\dots,n_\textrm{gr}$ runs over the frequency groups and $L(\nu < \nu_1)$ and $L(\nu > \nu_{n_\textrm{gr}})$ represent the contributes to the integral from frequencies outsides the extreme values of the groups. The terms can affect the results essentially only close to the upper temperature boundary of the opacity tables, where the blackbody function is not yet negligible at high frequencies. To extend the integral out of the group limits, we assume the opacity to be the same as the last available group, and explicitly calculate the other terms. We apply the same method every time an integration over the frequency space is required.

\subsection{Jet energy deposition}
\label{subsec:GRB_model}

If a GRB is launched from the polar regions of the remnant, it releases additional energy into the ejecta it traverses.
\knecnn\ can mimic this effect by adding an extra energy term to the right-hand side of Eq.~\eqref{eq:energy_cons} for the innermost shells in the polar regions of the ejecta.
We use the ``thermal bomb'' model already implemented in \snec\ \citep{Morozova:2015bla}.
The energy of the bomb is released during a chosen time interval and is exponentially attenuated both in time and with distance from the inner boundary.
We define two parametrized prescriptions for the energy released by the GRB.

In the ``simple jet'' model, we assume a jet with an opening angle $\theta_j$ to release a uniform (in $\theta$) isotropic energy $E_\text{iso} (\theta) = E_0$ in the traversed regions.
The total energy released by the jet is given by $E_j = (\Delta\Omega_j / 4\pi) E_0$, where $\Delta\Omega_j = 2\pi (1 - \cos\theta_j)$ is the solid angle covered by the jet (or two times this value if one assumes reflection with respect to the equatorial plane).

In the ``structured jet'' model, the bomb parameters are fixed assuming that the isotropic energy released from the GRB is proportional to the kinetic energy of the structured jet described by \cite{Ghirlanda:2018uyx}, given by $E_\text{iso} (\theta) = E_0 / [1 + (\theta/\theta_j)^{5.5}]$.
The total energy released by the jet can be obtained by dividing $E_\text{iso}(\theta)$ by the scaling factor $\lambda_\theta$ introduced in Sec.~\ref{subsec:rad-hydro}, and then integrating over the solid angle.
The energy release is taken to be constant over the duration $\Delta t_\text{jet}$ of the GRB emission, which starts at $t_\text{jet}$ after the start of the simulation.

\subsection{Initial profiles}
\label{subsec:init-cond}

Initializing \knecnn\ requires specifying an initial Lagrangian profile.
This profile must include the density $\rho$, temperature $T$, radial velocity $v$, entropy $s$ and electron fraction $Y_e$ as a function of the enclosed mass $m$ (including the mass of the remnant), and the position of the innermost fluid element $r_0$.
An estimate for the expansion timescale is also needed if the heating rate fits from \cite{Wu:2021ibi} are employed.
If the online NN is activated, the nuclear composition for each shell is initialized from the other thermodynamic quantities as described in Sec.~\ref{subsec:nnCoupling}.
From these initial data, \knecnn\ automatically defines a new profile with a prescribed number of mass shells $n_\text{sh}$ from a selected mass distribution function (i.e. the relative amount of mass $\delta m_i$ included in each shell $i=1,\dots,n_\text{sh}$, rescaled by the total mass of the ejecta $M_\text{ej}$) by performing a linear interpolation. The initial positions $r_i$ of the fluid elements are calculated  by progressively layering the mass shells as prescribed by the constraint $m(r) = 4\pi \int_{r_0}^{r_i} \rho(r) \, r^2 dr$ introduced in \cite{Wu:2021ibi}.

Our simulations are initialized with initial ejecta profiles extracted from ab-initio NR simulations performed with the \texttt{THC} code and including a microphysical EOS, M1 neutrino transport, and magnetic-field induced turbulence \citep{Radice:2018pdn, Perego:2019adq, Bernuzzi:2020tgt, Radice:2021jtw, Zappa:2022rpd, Bernuzzi:2024mfx}.
The main parameters of the considered binaries are listed in Table~\ref{tab:NR_params}.
All models have already been presented in
\cite{Bernuzzi:2024mfx,Gutierrez:2024pch}.
For details on the EOS used, see \cite{Hempel:2009mc, Typel:2009sy, Bombaci:2018ksa, Logoteta:2020yxf}.
Both selected models \texttt{BLh\_1.43} and \texttt{DD2\_1.67} describe asymmetric BNSM.
The chirp mass of the \texttt{BLh\_1.43} configuration is compatible with GW170817, and the mass ratio lies at the upper bound of the constraints inferred with low spin priors \cite{Abbott:2018wiz}.
The \texttt{DD2\_1.67} represent a comparable but more extreme mass-ratio scenario.
The neutron stars composing the binary systems are all nonrotating.
The central remnant is a massive neutron star for the duration of the simulation in both models.

Throughout each simulation, the unbound material is identified via the Bernoulli criterion \citep[\eg][]{Nedora:2020hxc} and collected at an extraction radius $r_\text{ext} \simeq 443$~km.
Both models eject matter for the full duration of the simulations until $\sim100$~ms post merger.
In order to prepare the subsequent long-term evolution, the ejecta properties are averaged over the azimuthal coordinate, while the polar dependence is retained by discretizing the polar angle into $51$ uniformly spaced angular sections.
The entropy is recalculated from the original EOS to avoid averaging artifacts and ensure thermodynamic consistency.
The extracted time-dependent ejecta profile is then mapped into a Lagrangian profile by positioning the latest shell at $r_0 = r_\text{ext}$, and progressively layering previously ejected shells on top.

In constructing the Lagrangian ejecta profiles, we neglect non-radial components of the ejecta velocity.
Moreover, owing to the ray-by-ray nature of the simulations, possible fallback at later times is largely suppressed, since outer shells can only fallback if the inner shells are also infalling.
We do not expect these assumptions to introduce strong qualitative corrections to the results discussed in this paper, and defer a quantitative assessment of their impact on the ejecta evolution and final observables to future work.

\begin{figure*}
	\centering 
	\hspace*{-5pt}\includegraphics[width=\textwidth]{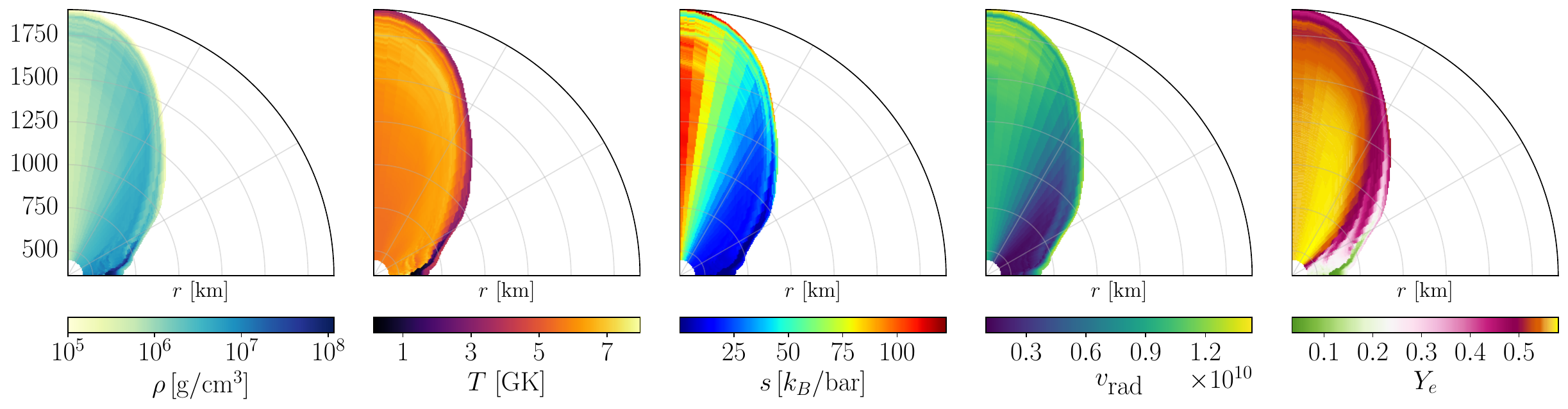}
	\caption{Thermodynamic profile of the ejecta at the beginning of the \knecnn\ simulation for the \texttt{BLh\_1.43} binary. From left to right, the quadrants show the initial density, temperature, entropy, radial velocity, and electron fraction.}
	\label{fig:initial_profiles}
\end{figure*}

The initial profile extracted from the \texttt{BLh\_1.43} simulation is shown in Fig.~\ref{fig:initial_profiles}.
This model produces a massive, neutron-rich ($Y_{e,0} \lesssim 0.22$) tidal tail along the equatorial regions.
Emissions from the long-lived remnant drive a proton-rich ($Y_{e,0} > 0.5$) neutrino-driven wind in the polar regions above the disk persisting until the end of the simulations.
The \DD2\ case is qualitatively similar, but the higher mass ratio strongly enhances the tidal-tail dynamical ejection, while less matter is ejected in the polar neutrino-driven wind.
In what follows, we mainly focus on the \BLh\ profile to systematically asses the impact of refined nuclear burning and atomic opacities models on the ejecta dynamics and nucleosynthesis.
The \DD2\ case is also considered to verify the robustness of our results.
\begin{table*}
\centering
\caption{Main parameters of the ab-initio NR simulations from which we generate the initial profiles for the \knecnn\ runs discussed in this paper.}
\begin{tabular}{ccccccccc}
	\hline
	Name & EOS & $q$ & $M_1 [M_\odot]$ & $M_2 [M_\odot]$ &
		$t_f [\text{ms}]$ & $t_\text{dyn} [\text{ms}]$ & $M_\text{dyn} [10^{-2} M_\odot]$ &  $M_\text{wind} [10^{-2} M_\odot]$ \\
	\hline
	\texttt{BLh\_1.43} & BLh & 1.43 & 1.146 & 1.635 & 103.5 & 34.1 & 0.256 & 0.827 \\
	\texttt{DD2\_1.67} & DD2 & 1.67 & 1.80 & 1.08 & 112.8 & 37.5 & 1.964 & 0.567 \\
		\hline
\end{tabular}
\tablefoot{From left to right, the name of the simulation, the EOS employed, the mass ratio $q$, the gravitational masses of the two non-rotating neutron stars, $M_1$ and $M_2$,
		the final time $t_f$ of the simulation, the time we choose to distinguish between dynamical and wind ejecta $t_\text{dyn}$, and the mass of these two components, $M_\text{dyn}$ and $M_\text{wind}$. The times are reported here as times after merger.
	The remnants do not collapse to a black hole before the simulations end.}
\label{tab:NR_params}
\end{table*}

\section{Results}
\label{sec:results}

We analyze a series of simulations performed on the \BLh\ and \DD2\ ejecta profiles to explore the physical impact on the nucleosynthesis and light-curve calculations from the introduction of the in-situ NN, as well as the effects of switching between the \texttt{ST05} and \texttt{DT08} thermalization schemes or among the opacity models described in Sec.~\ref{subsec:opacity}.
The simulations considered in this section and their respective configurations are listed in Table~\ref{tab:knecnn_runs}.
\begin{table*}[t]
	\centering
	\caption{List of \knecnn\ simulations analyzed in this paper. 	}
		\begin{tabular}{cccccccccc}
		\hline
		Name & NR profile & NSE & Therm & $\kappa(\nu)$ & Opacity & Atomic data & $L_\text{peak}$ [$10^{41}$ erg/s] & $t_\mathrm{peak}$ [hours] & $X_{\text{lan},f}$ [$\%$] \\
		\hline
		\texttt{NdU300} & \texttt{BLh\_1.43} & \texttt{BK8} & \texttt{DT08} & $\kappa_{\nu,300}$ & LANL & Nd, U & 6.404 & 1.12 & $2.400$ \\
				\texttt{Nd300} & \texttt{BLh\_1.43} & \texttt{BK8} & \texttt{DT08} & $\kappa_{\nu,300}$ & LANL & Nd & 5.668 & $1.00$ & $2.400$ \\
		\texttt{GK} & \texttt{BLh\_1.43} & \texttt{BK8} & \texttt{DT08} & Ross. & LANL & Nd & $7.511$ & $2.168$ & $2.401$ \\
		\texttt{GKPl} & \texttt{BLh\_1.43} & \texttt{BK8} & \texttt{DT08} & Planck & LANL & Nd & $3.166$ & $0.84$ & $2.401$ \\
		\texttt{JK} & \texttt{BLh\_1.43} & \texttt{BK8} & \texttt{DT08} & Ross. & J[22] & - & $3.495$ & $0.83$ & $2.401$ \\
		\texttt{SK} & \texttt{BLh\_1.43} & \texttt{BK8} & \texttt{DT08} & Ross. & W[22] & - & $1.663$ & $0.75$ & $2.400$ \\
				\texttt{BK6} & \texttt{BLh\_1.43} & \texttt{BK6} & \texttt{DT08} & Ross. & LANL & Nd & $6.878$ & $2.23$ & $2.498$ \\
		\texttt{CNSE} & \texttt{BLh\_1.43} & \texttt{CNSE} & \texttt{DT08} & Ross. & LANL & Nd & $7.315$ & $2.00$ & $2.389$ \\
		\texttt{Th-S} & \texttt{BLh\_1.43} & \texttt{BK8} & \texttt{ST05} & Ross. & LANL & Nd & 4.581 & 1.50 & 2.507 \\
		\texttt{ThK-S\_BLh} & \texttt{BLh\_1.43} & \texttt{BK8} & \texttt{ST05} & Ross. & W[22] & - & $1.247$ & $1.45$ & $2.507$ \\
		\texttt{Apr2\_BLh} & \texttt{BLh\_1.43} & - & \texttt{ST05} & Ross. & W[22] & - & $4.672$ & $0.94$ & $2.837$ \\
		\hline
				\texttt{ThK-S\_DD2} & \texttt{DD2\_1.67} & \texttt{BK8} & \texttt{ST05} & Ross. & W[22] & - & $1.870$ & $3.00$ & $5.546$ \\
		\texttt{Apr2\_DD2} & \texttt{DD2\_1.67} & - & \texttt{ST05} & Ross. & W[22] & - & $4.536$ & $2.17$ & $6.319$ \\
						\hline
	\end{tabular}
	\tablefoot{From left to right, the name of the simulation and of the NR-extracted initial ejecta profile (see Table~\ref{tab:NR_params}), 		the NSE initialization method employed (\texttt{BK6} and \texttt{BK8} stand for the backtracking procedure starting at $T=6,8$~GK, \texttt{CNSE} for the cold-NSE method; \texttt{Apr2} simulations are performed without in-situ NN), the thermalization scheme used (\texttt{DT08} and \texttt{ST05} stand for the detailed thermalization scheme described in Sec.~\ref{subsec:thermalization} with $f_\text{th}^\text{oth}=0.8$ or the simple thermalization from \cite{Wu:2021ibi} with $f_\text{th}^\text{\knec}=0.5$).
		The next columns state wether a frequency-dependent (and with how many frequency groups) or a gray opacity is used, which model is employed (LANL stands for the atomic calculations, J[22] for the analytical model from \cite{Just:2021vzy}, and W[22] for the opacity from \cite{Wu:2021ibi}), and which atomic opacities are read from the NIST-LANL database (see Sec.~\ref{subsec:opacity}).
		The last three columns report the peak of the bolometric luminosity, the time of this peak, and the final mass fraction of lanthanides produced.
							}
	\label{tab:knecnn_runs}
\end{table*}

\subsection{Effects of dynamics on nucleosynthesis}
\label{subsec:res_onlineNN}

The online nuclear calculations implemented in \knecnn\ allow for a coupled evolution between the ejecta hydrodynamics and its composition.
In \cite{Magistrelli:2024zmk}, we showed that prescribing the density evolution with the analytical model of \cite{Lippuner:2015gwa} can lead to inaccurate nucleosynthesis predictions.
Capturing the detailed early ($\lesssim 500$~ms) dynamics of the outflow is essential for reliably determining the nuclear yields and the abundance evolution across the entire nuclide chart.
On the other hand, the self-consistently computed nuclear heating can, in principle, feed back on the ejecta evolution.
Access to the instantaneous isotopic composition is also required to include detailed physical models for the thermalization of the released nuclear power and the optical opacity.
In this and the next sections, we quantify the influence of the dynamics on nuclear calculations, and assess the impact of the improved nuclear power alone on the ejecta dynamics and on the resulting light curves.
We deliberately exclude here the composition-dependent thermalization and opacity models (made possible only by the NN coupling), and defer the discussion of their effects to Sec.~\ref{subsec:res_thermalization} and \ref{subsec:res_opacity}.
We therefore only focus in this section on the simulations labeled \texttt{ThK-S} and \texttt{Apr2} in Table~\ref{tab:knecnn_runs}, both for the \BLh\ and \DD2\ ejecta profiles. 

For the runs without a coupled NN, we compute the nucleosynthesis using two different post-processing procedures.
In the first approach (\texttt{pp\_grid}), following the same method explored in \cite{Magistrelli:2024zmk}, we take the $Y_e, s, \tau$ grid defined in \cite{Perego:2020evn} and add one dimension for the temperature.
We include 15 linearly spaced points between the minimum and maximum temperatures recorded in the initial \knecnn\ profile.
To account for proton-rich winds, we also extend the $Y_e$ dimension with 11 additional points uniformly distributed between $0.5 \le Y_e \le 0.7$. We then map all of the initial fluid elements onto this grid and run an independent instance of \skynet\ for each of the grid points, prescribing an exponential-plus-homologous expansion as in \cite{Lippuner:2015gwa}, and defined from the initial ejecta profile as in \cite{Perego:2020evn}.
The NSE initialization employs the same backtracking prescription used in the full \knecnn\ run.
The final results are obtained by averaging the isotopic abundances over the grid, weighting each point by the total mass of the represented fluid elements.
In the second post-processing method (\texttt{pp\_traj}), we instead evolve a NN along the thermodynamic trajectory of each individual fluid element from the original \knecnn\ simulation, and then again mass-weight the abundances.

Comparing the \texttt{pp\_grid} procedure against the online results highlights the effects of the detailed early ($\lesssim 500$~ms) dynamics on the nucleosynthesis.
During this phase, the expansion of the outflow is shaped by the details of the initial explosion and is dynamically influenced by the released nuclear energy as well as by interactions among fluid elements.
Comparing instead the \texttt{pp\_traj} method against the online results allows us to isolate the impact of self-consistent nuclear-energy feedback on the dynamics, and consequently again on the nuclear calculations, relative to a simpler analytical prescription for the nuclear power.

\begin{figure*}
	\centering 
	\includegraphics[width=\textwidth]{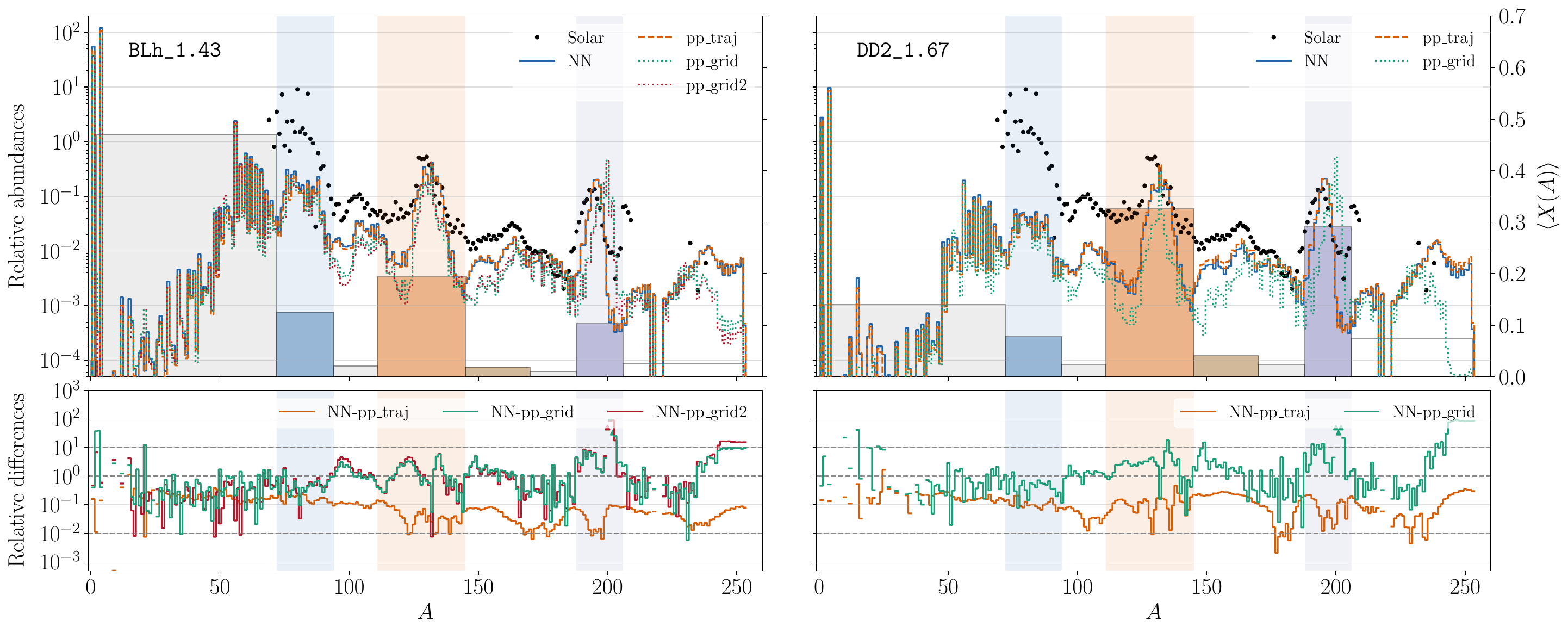}
	\caption{Mass-weighted nucleosynthesis yields at $t \simeq 5 \times 10^4$~s for the \BLh\ (left column) and \DD2\ (right column) ejecta profiles.
		\textit{Top panels}: Final yields for the simulations performed with in-situ NN (blue) or using analytical fits for the nuclear power and post-processing the nucleosynthesis either on the original thermodynamic trajectories (orange), or on analytical density evolutions prescribed from a grid of initial thermodynamic conditions, with a standard (green) or double (red) resolution on the $Y_e, \tau, s$ grid.
		The black dots represent the Solar $r$-process abundances \citep{Prantzos:2020a}.
		All the abundances are scaled to produce a unitary total abundance of the elements with $170 \leq A \leq 200$.
		The histogram shows, for the in-situ NN run, the global cumulative mass fractions of the first (blue), second (orange) and third (purple) $r$-process peaks, and the rare-earths (brown).
				\textit{Bottom panel}: relative differences against the simulation with in-situ NN. The upward [downward] triangles indicate discrepancies greater than two orders of magnitude [smaller than $5\times10^{-4}$]. The horizontal dashed lines highlight the $1\%$ and factors of 1 and 10 discrepancies.}
	\label{fig:yields_onlineNN}
\end{figure*}

We first investigate how the detailed hydrodynamic evolution impacts the nuclear calculations.
In Fig.~\ref{fig:yields_onlineNN}, we compare the final nucleosynthesis yields obtained with in-situ NN against those predicted using nuclear-power fits and the two post-processing methods.
We also test the \texttt{pp\_grid} method by doubling the number of points in the $Y_e, s, \tau$ dimensions to assess the impact of grid resolution on the final abundances.
The two asymmetric simulations give similar nucleosynthesis, whose general features are confirmed between among the different procedures.
Strong $r$-process reactions dominate the equatorial, low-$Y_e$ regions of the tidal tails.
The light ($A \lesssim 110$) element peaks, mainly produced in the ejecta emitted above the remnant disk ($\theta \gtrsim 30$~degrees), are more pronounced in the \BLh\ profile, which features a less prominent tidal tail.

The \texttt{pp\_grid} procedure leads to inaccurate final nuclear yields (even when the effect of the released nuclear power is included in the temperature evolution).
Relative differences of $\gtrsim 1$ appear across the entire range of mass numbers, with deviations approaching an order of magnitude for several isotopes within the second and third $r$-process peaks.
A noticeable shift of the third peak is also evident when comparing these results to those from the simulation employing in-situ NN.
The grid discretization has only a secondary effect, as doubling the resolution on the $Y_e, \tau, s$ dimensions does not significantly affect the results. %

The \texttt{pp\_traj} analysis reproduces most of the final abundances within a $\sim 20\%$ error.
This improvement comes from the more accurate representation of the density (and temperature) evolution along each trajectory. If (opacity and) thermalization effects are neglected, and if reliable analytical fits for the nuclear powers are employed, a sufficiently large sample of tracers, well resolved in time up to $\sim1$~s, can therefore yield reliable nucleosynthesis predictions (if non-radial motion is neglected).
The remaining discrepancies originate from corrections to the thermodynamic evolution of the fluid elements that come exclusively from the improved nuclear-power calculation provided by the in-situ NN (see next section for details).
The effect on the final mass fraction of lanthanides is of the order of $10\%$, as reported in Table~\ref{tab:knecnn_runs}.

\begin{figure*}
	\centering 
	\hspace*{-10pt}\includegraphics[width=\textwidth]{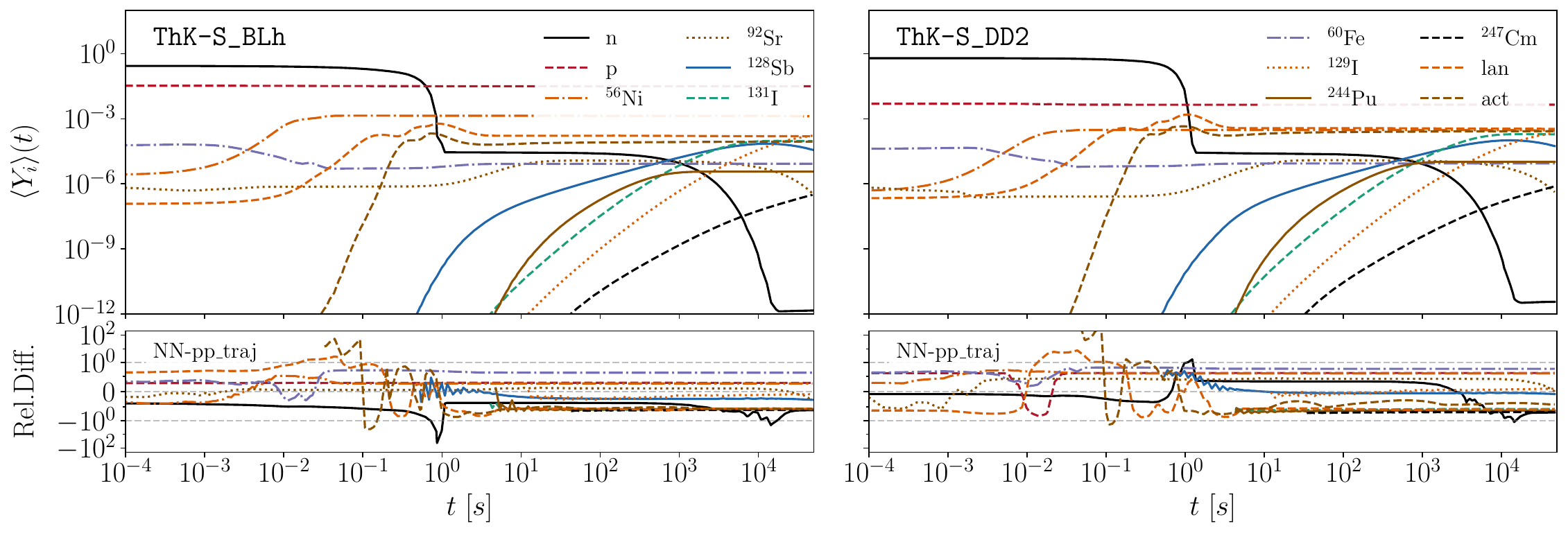}
	\caption{Time evolution of the mass-weighted abundances for selected isotopes and cumulative abundances of lanthanides and actinides.
		The first and second columns refer to simulations \texttt{ThK-S\_BLh} and \texttt{ThK-S\_DD2}, respectively.
	The top and bottom rows show the results obtained with in-situ NN and their relative differences with respect to the post-processed abundances computed with the tracer method.
	The horizontal dashed lines in the bottom panels indicate factors-of-unity deviations and the case $\langle Y_i \rangle = \langle Y_i^{pp} \rangle$.
	Differences are displayed only at times for which $\langle Y_i \rangle > 10^{-12}$.
		}
	\label{fig:spec_abs_NN}
\end{figure*}

In Fig.~\ref{fig:spec_abs_NN}, we compare the time evolution of selected species obtained from the runs with in-situ NN to those from the post-processing tracers procedure.
We focus on neutrons and protons as diagnostic isotopes, $^{56}$Ni (half-life $\tau \simeq 6.077$~days), $^{92}$Sr (with $\tau \simeq 2.611$~hr), $^{128}$Sb (with $\tau \simeq 9.05$~hr) and $^{131}$I (with $\tau \simeq 8.02$~days) for their potential contributions to $\gamma$-ray emission \citep{Korobkin:2019uxw, Bernuzzi:2024mfx, Jacobi:2025eak}, $^{60}$Fe, $^{129}$I, $^{244}$Pu and $^{247}$Cm for their relevance in galactic chemical evolution studies \citep{Cote:2021zkj, Wallner:2021, Davis:2022, Chiesa:2023jno}, and the cumulative mass fractions of lanthanides and actinides to monitor strong $r$-process nucleosynthesis.
At $t\sim1$~s, the system departs from $(n,\gamma) \leftrightarrow (\gamma,n)$ equilibrium.
The neutron-to-seed ratio consequently drops, triggering neutron freeze-out.
The remaining free neutrons undergo $\beta$-decay on the $t\sim10$~minutes timescale.
$^{56}$Ni and lanthanides are already present at $t=0$.
$^{56}$Ni is further synthesized on the $t\sim10$~ms timescale, while lanthanides are produced for $t\gtrsim100$~ms together with actinides.
The isotopes $^{128}$Sb, $^{129}$I, and $^{131}$I are produced at significant abundances on a timescale of $t\sim1$~hour, with $^{128}$Sb already decaying at $t\sim10$~hours.

The neutron freeze-out occurs earlier in the online calculations than in the post-process predictions for the \BLh\ profile, and later in the \DD2\ case.
In the two cases, the post-processing approach overestimates or underestimates, respectively, the amount of remaining free neutrons by $\sim10\%$.
The final neutron abundances agree within a few tens of percent, with the online NN consuming more neutrons during the last stages of the evolution.
For both profiles, lanthanides and actinides are overproduced by the post-processing calculations by a few tens of percent.
The same level of discrepancy represents an approximate upper limit for the other selected elements, but significantly larger deviations (up to two or more orders of magnitude) can be observed at early times ($t \lesssim 1$~s).
The abundance evolution of $^{129}$I seems unaffected by the online calculations.

\subsection{Effects of nucleosynthesis on dynamics and light curves}
\label{subsec:res_onlineNN_lc}

\begin{figure*}
	\centering 
	\includegraphics[width=\textwidth]{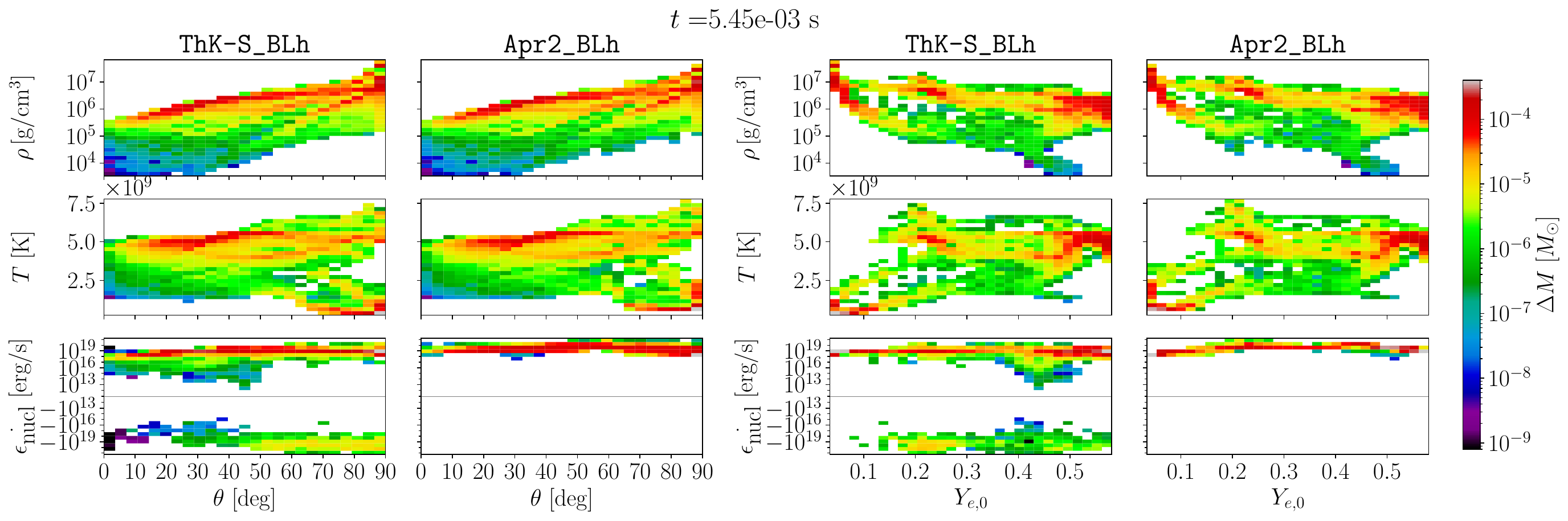}\\
	\vspace*{10pt}
	\includegraphics[width=\textwidth]{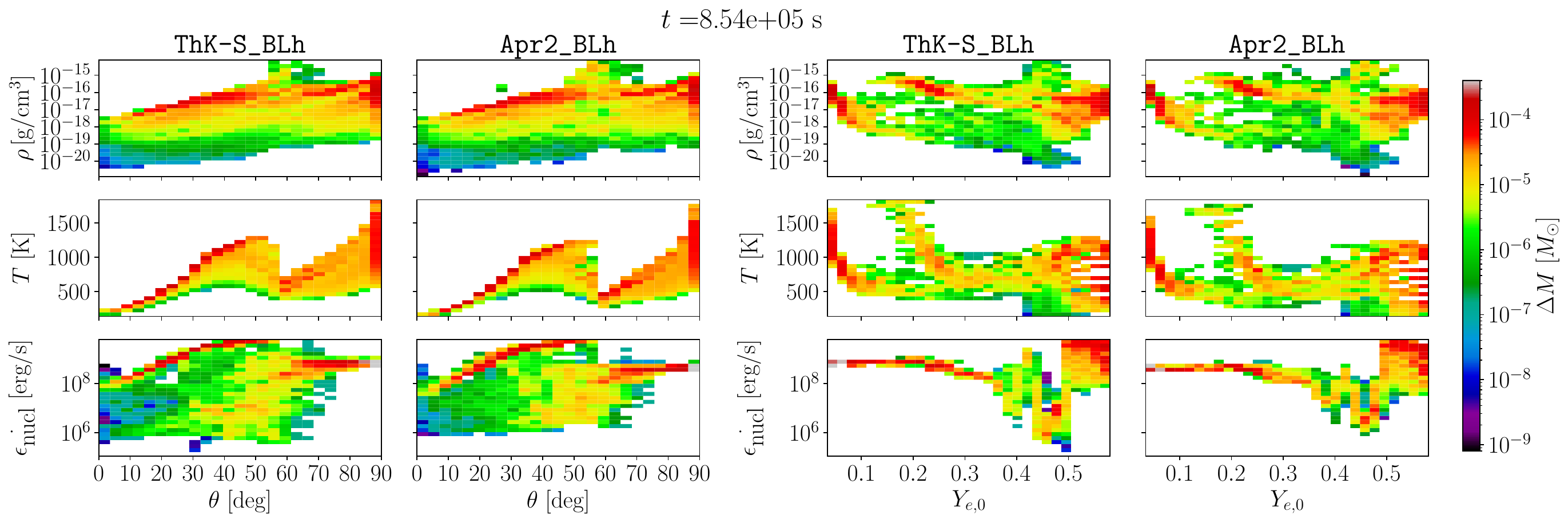}
	\caption{Mass-weighted histograms of the density, temperature and instantaneous thermalized heating rate obtained from the \BLh\ runs using the in-situ NN (\texttt{ThK-S\_BLh}) or the analytical fits for the nuclear power (\texttt{Apr2\_BLh}). 		The top and bottom groups of plots show the ejecta at early (pre neutron freeze-out) and late ($t\simeq10$~days) times.
		The left [right] column shows the profiles as functions of the polar angle [initial electron fraction]. The gray lines in the heating rate panels separate negative and positive values.}
	\label{fig:hydro_onlineNN}
\end{figure*}

We check the impact of the in-situ calculation of the nuclear power on the ejecta hydrodynamics by analyzing the density, temperature, and thermalized heating rate evolution using the histograms in Fig.~\ref{fig:hydro_onlineNN}.
At early times, the thermodynamic evolution of the outflow is dominated by the initial explosion and is only weakly affected by nuclear burning.
At later times, nuclear heating becomes the main source of energy in the ejecta, increasing the temperature of material with initial electron fraction $Y_{e,0}\lesssim0.1$ by a factor of two or more due to the decay of freshly synthesized neutron-rich isotopes.
For $t\gtrsim1$~day, nuclear heating also raises the temperature in regions with $Y_{e,0}\gtrsim0.45$, when $^{56}$Ni begins to undergo $\beta$-decay.

The analytical fits fail to reproduce the detailed structure and spread of the nuclear power distribution (especially at early times) and tend to cluster fluid elements around the regions already highly populated in the in-situ NN case.
Nevertheless, they globally recover the correct orders of magnitudes of the heating rates.
Consequently, the impact on the hydrodynamic evolution is limited, yielding only minor deviations in the density and temperature profiles over time. As a result, the hydrodynamics corrections introduced by the online NN have a negligible effect on the nucleosynthesis predictions.

\begin{figure*}
	\centering 
	\hspace*{-5pt}\includegraphics[width=\textwidth]{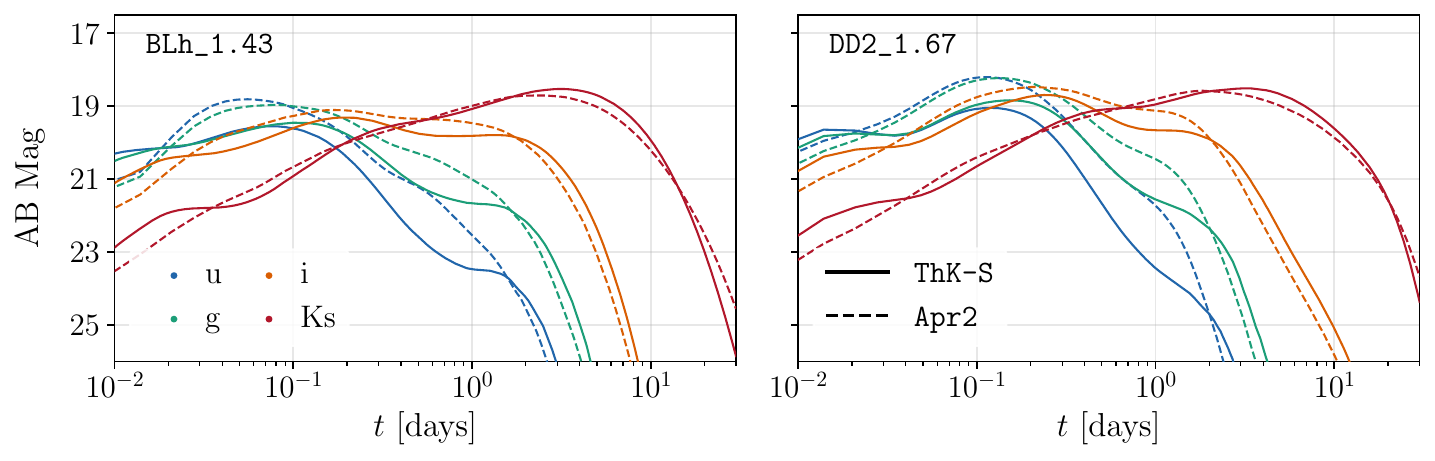}
	\caption{Predicted AB apparent magnitudes in the Gemini $u$, $g$, $i$ and $K_s$ bands for an observer at a polar angle of $30$~degrees and a distance of 40~Mpc.
		Comparison between the simulations using in-situ NN (solid lines) and analytical nuclear power fits (dashed lines). 		All models adopt simple (no composition-dependent) thermalization and opacity models.
		The left and right panels show results for the \BLh\ and \DD2\ ejecta profiles, respectively.}
	\label{fig:light_curves_NN}
\end{figure*}

The differences in the predicted evolution of the photosphere temperature and the nuclear power released in optically thin regions, however, have a qualitative impact on the light curves.
In Fig.~\ref{fig:light_curves_NN}, we compare kilonova predictions computed with and without in-situ NN for the \BLh\ and \DD2\ ejecta profiles.
The light curves are brighter, peak later, and last longer in the \DD2\ case, which produces more than double the amount of ejected mass, as expected from the Arnett's law \citep{Arnett:1982}.
Both simulations show a blue peak at $t\sim1$~hour, when the photosphere enters regions previously heated up by the decay of freshly synthesized $r$-process isotopes \citep{Bernuzzi:2024mfx}.
The late-time red kilonova is powered both by the ongoing decays of these neutron-rich isotopes, and by the $^{56}$Ni $\to ^{56}$Co $\to ^{56}$Fe decay chain \citep{Jacobi:2025eak}.

\begin{figure}
	\centering
	\hspace*{-5pt}\includegraphics[width=0.5\textwidth]{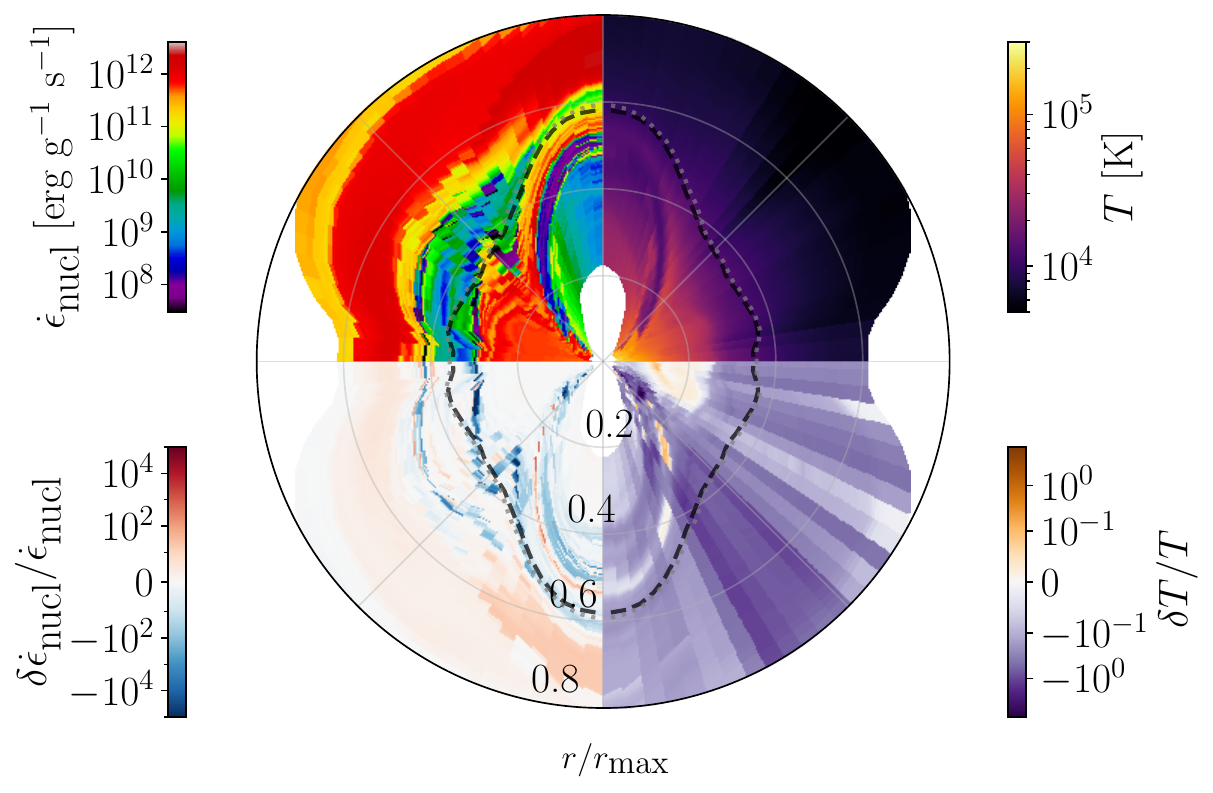}
		\caption{Thermalized heating rate and temperature profiles of the \BLh\ ejecta at $t\sim1$~hour.
		The top panels show the results obtained with in-situ NN, and the bottom panels show the relative differences with respect to the run using analytical nuclear power fits.
		The dashed black and the dotted gray lines represent the position of the photosphere in the runs with in-situ NN and fits, respectively.}
	\label{fig:hydro_tempHeat2D}
\end{figure}

Calculating the nuclear power on the fly generally shifts the emission peak to later times and lower frequencies (see also Table~\ref{tab:knecnn_runs}).
The lower luminosity at $t \sim 1$~hour is caused by the colder temperature predicted by the online NN for the photosphere, which most contributes to the light curves on this timescale.
We show this effect in Fig.~\ref{fig:hydro_tempHeat2D}, which displays the heating rate and temperature profiles for the \BLh\ simulation at $t\sim1$~hour.
At this time, nuclear contributions from the optically thin layers of the ejecta provide only a minor correction to the luminosity, and differences between the in-situ and ex-situ simulations arise only in low-activity regions.
Similar considerations hold for the \DD2\ profile.

The plateau in the blue, green, and yellow filters at $t\sim1$~day is sustained by the decays of $^{56}$Ni and, more generally, by the reactions occurring in the initially mildly neutron-rich and proton-rich fluid elements.
Their effect is overestimated by the simulations relying on the analytical nuclear power fits.
At later times, the decays of $r$-process isotopes also contribute significantly to the EM emission.
The online NN slightly enhances the light curves across all frequencies with respect to the offline calculations, but predicts fainter red kilonovae at very late times ($t\gtrsim10$~days).
The dimmer luminosity at $t\sim10$~days is consistent with the heating rates presented in the bottom part of Fig.~\ref{fig:hydro_onlineNN}, where the offline calculations are shown to slightly underestimate heating rates all across the ejecta.
Our comparative analysis is not affected by non-LTE effects, which become relevant at these late times, when the ejecta enters the nebular phase \citep[\eg][]{Pognan:2022pix, Ricigliano:2025knd}. %

While the light curves appear reasonably accurate at late times in this stage of the analysis, tracking the ejecta composition is needed for the detailed calculations of the improved thermalization factors in the \texttt{DT08} scheme and for the \texttt{JK}, \texttt{LANL-R}, \texttt{LANL-p}, and \texttt{LANL} opacity models.
Their relevant qualitative impacts on the light curve predictions are discussed separately in Sec.~\ref{subsec:res_thermalization}~and~\ref{subsec:res_opacity}, while the overall improvements enabled by the online NN are discussed in Sec.~\ref{subsec:res_best_comparison}.

\subsection{Effects of thermalization treatment}
\label{subsec:res_thermalization}

The thermalization prescription directly enters the nuclear term in Eq.~\eqref{eq:energy_cons}.
However, nuclear heating remains a subdominant contribution to the dynamics for most of the ejecta evolution.
We show this behavior in Fig~\ref{fig:check_cons}, which displays the instantaneous average contributions to the internal energy of the fluid for the \BLh\ profile under different thermalization and opacity prescriptions.
On the timescales relevant for the nucleosynthesis, the effects of the different thermalization schemes on the temperature and density evolution are almost negligible.
As a result, for the \BLh\ profile, the final nuclear yields agree within roughly $\sim10\%$, and the final mass fraction of lanthanides is consistent to within a few percent (see Table~\ref{tab:knecnn_runs}).
The isotopic evolution is similar for both thermalization schemes.
Neutron freeze-out occurs slightly earlier (by $\sim200$~ms) in the simple thermalization case, with no qualitative change in the rest of the isotopic evolution.
\begin{figure}
	\centering
	\hspace*{-5pt}\includegraphics[width=0.5\textwidth]{fig08.pdf}
	\caption{
	Contributions to the internal energy of the \BLh\ ejecta for the simple and detailed thermalization schemes described in Sec.~\ref{subsec:thermalization} (\textit{left column}), and for selected opacity models from Sec.~\ref{subsec:opacity} (\textit{right column}).
	\textit{Top:} Average pressure work (blue), nuclear heating (orange), and radiation energy (red) per unit time and per fluid element, mass-weighted over the ejecta.
	\textit{Bottom:} Ratio of nuclear heating and radiation energy over pressure work.
	The horizontal dashed line marks a $30\%$ ratio.
		}
	\label{fig:check_cons}
\end{figure}

\begin{figure}
	\centering 
	\hspace*{-12pt}\includegraphics[width=0.5\textwidth]{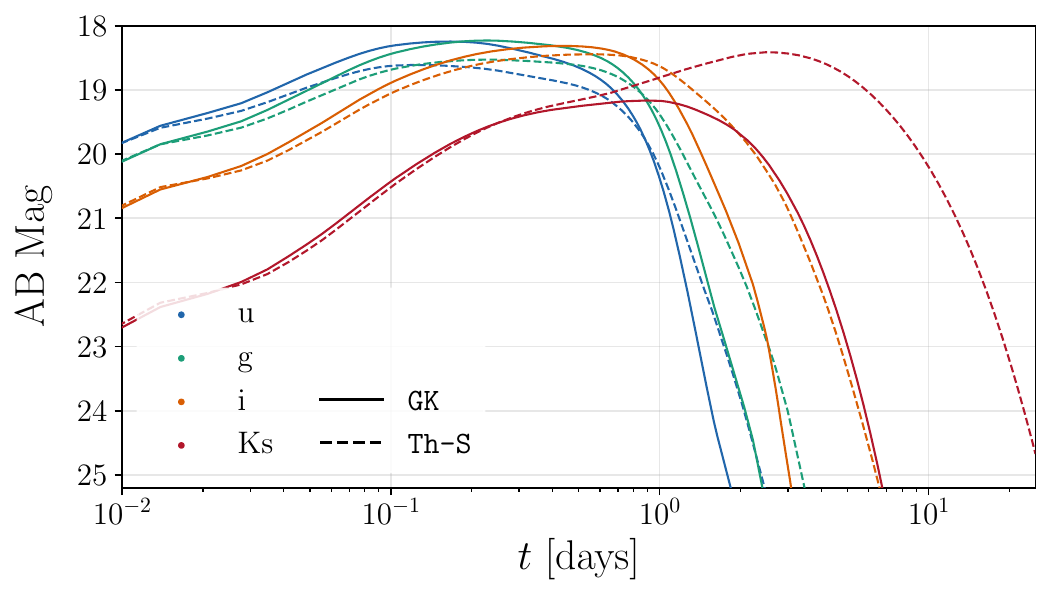}
	\caption{Predicted AB apparent magnitudes in the Gemini $u$, $g$, $i$ and $K_s$ bands for an observer at a polar angle of $30$~degrees and a distance of 40~Mpc.
		Comparison between the \BLh\ simulations using in-situ NN and either the time-dependent, composition-based thermalization scheme (solid lines) or a constant thermalization factor (dashed lines).}
	\label{fig:light_curves_therm}
\end{figure}

At $t \gtrsim 1$~day, the simple \texttt{ST05} thermalization scheme significantly overestimates the fraction of nuclear power that is actually thermalized and thus contributes to the temperature and EM emission of the ejecta.
This result is independent of the specific choice of the constant $f_\text{th}^\text{\knec}$, as the thermalization efficiency of all decay products drops exponentially on the days timescales.
At these times, the outflow begins to become optically thin, and the thermalized energy is promptly emitted as thermal radiation from the layers above the photosphere.
As shown in Fig~\ref{fig:check_cons}, the radiative power is predominantly supplied by nuclear heating at these timescales, while the photospheric contribution becomes negligible.
The effect on the light curves is evident from Fig.~\ref{fig:light_curves_therm}, which displays the kilonova predicted by the two thermalization schemes.
The simple \texttt{ST05} prescription boosts the luminosity in all filters at late times compared to \texttt{DT08}, giving a very long-lived kilonova.
The effect is particularly pronounced in the low-frequency bands, as the ejecta has cooled enough at these times ($T\sim2000$~K) to have its blackbody spectrum peaking in the infrared.

At early times, the thermalization efficiency is close to unity for all decay products.
Any constant, fractional thermalization factor therefore underestimates the nuclear contribution to the internal energy of the ejecta.
The resulting lower temperature of the photosphere explains the dimmer and redder light curves produced by the simple thermalization scheme at $t\!\sim$~hours and the lower bolometric luminosity reported in Table~\ref{tab:knecnn_runs}.
The two models cross between roughly $\sim\!6$~hours and $\sim\!1$~day, when the effective averaged thermalization factor from the online calculations drops to $\langle f_\text{th} \rangle \sim0.5$.

\begin{figure}
	\centering 
	\hspace*{-12pt}\includegraphics[width=0.5\textwidth]{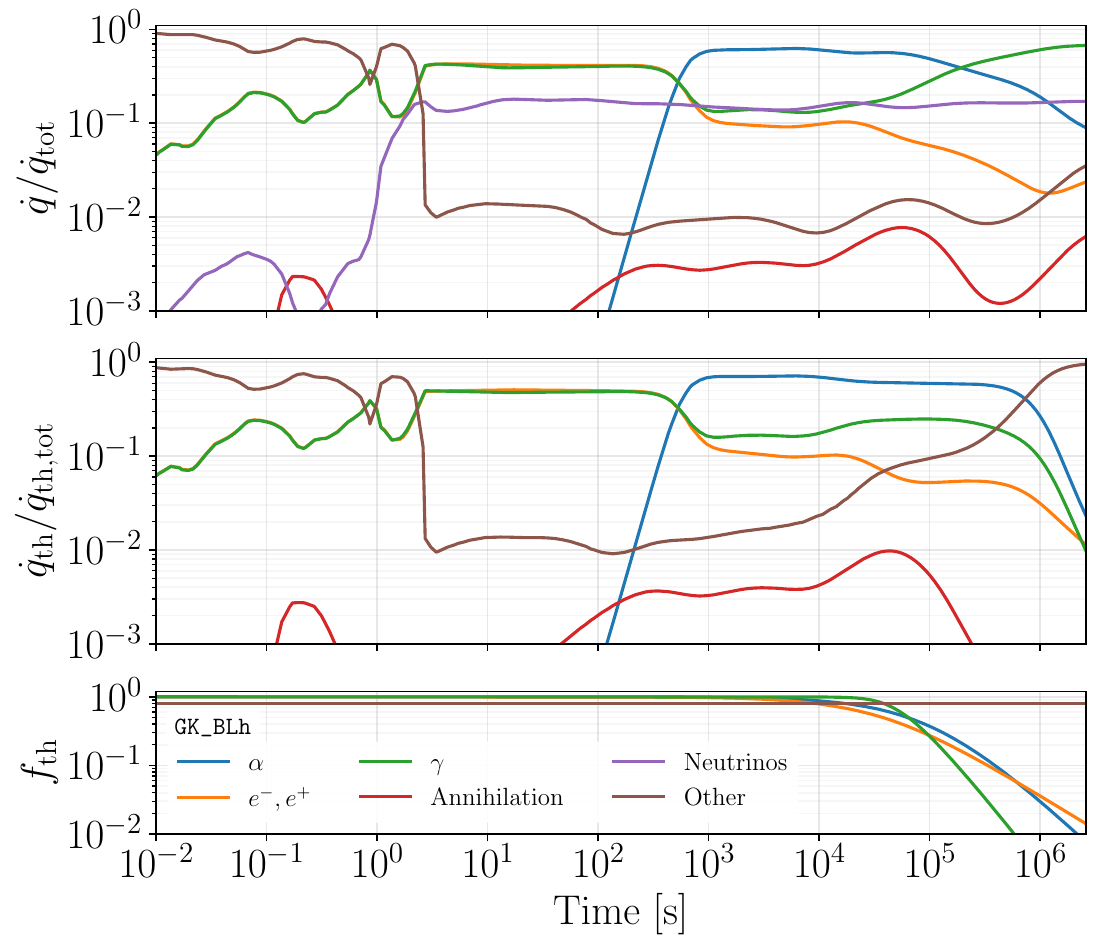}
	\caption{Mass-weighted contributions to the nuclear power (\textit{top panel}) and heating rates (\textit{central panel}) from $\alpha$ particles, electrons and positrons, $\gamma$-rays (direct emission and $e^-/e^+$ annihilation), neutrinos, and all remaining sources of injected energy, and their thermalization efficiency (\textit{bottom panel}) over time for the \texttt{GK} simulation.}
	\label{fig:thermalization}
\end{figure}

Figure~\ref{fig:thermalization} shows the mass-weighted contributions to the nuclear power and thermalized heating rate from the species considered in Sec.~\ref{subsec:thermalization}, and their thermalization efficiency over time as calculated by the \texttt{DT08} model.
At the $t\lesssim1$~s timescale, the nuclear power is dominated by the residual term.
In the neutron-rich parts of the outflow, contributions to this term come from neutron emissions from the $(\gamma,n) \leftrightarrow (n,\gamma)$ equilibrium and $\beta$ decays of very neutron-rich isotopes not included in the \ENDF\ database.
In the proton-rich regions, proton emissions contribute to the residual nuclear power on the other side of the valley of stability.
The same considerations hold for the heating rate, as all injected particles thermalize efficiently at very early times.
At $1~\textrm{s} \lesssim t \lesssim 1$~hour, the nuclear power is approximately equally injected in electrons, positrons, prompt $\gamma$-rays and partially neutrinos.
The latter are removed from the heating rates, while the hierarchy of the other contributions is influenced by the heating rates from $t\gtrsim1$~hour, when the specific thermalization factors begin to show qualitatively different behaviors.
At $t\gtrsim1$~hour, $\alpha$ decays also become relevant.
Their contribution is partially overestimated, as some $\alpha$ emitters are not correctly included in the REACLIB database used by \skynet, but they are accounted in the \ENDF\ database.
The effect is a boost of the relative fraction of energy associated with $\alpha$ particles (the total nuclear power is not affected, as it is directly obtained as an output from the NN). The effects on the total heating rates are expected to be minor, as while an overestimated fraction of the nuclear power will be thermalized as $\alpha$ particles, their thermalization efficiency drops exponentially in time.

At $t \gtrsim 1$~day, the residual term begins again to be comparable with the other sources of thermalized heating rates.
Only part of this behavior is due to physical contributions from the recoil energy of daughter nuclei and fission fragments, which are expected to always thermalized efficiently.
Part of the late-time residual nuclear power is due to small discrepancies between the nuclear inputs in \skynet\ and the \ENDF\ database.
This component becomes relevant in the heating rate because of its constant thermalization efficiency ($f_\text{th}^\text{oth} = 0.8$ in our simulation), whereas the other energy contributions are either strongly suppressed by thermalization factors that are $\lesssim 5\times10^{-2}$ for charged particles and $\lesssim 5\times10^{-3}$ for $\gamma$-rays, or lost entirely (\eg\ neutrinos).
If a small fraction of these suppressed contributions is incorrectly reassigned to the residual heating channel, it will thermalize too efficiently and consequently lead to an overestimate of the late-time luminosity.
The same limitations exist in the simpler thermalization method of \cite{Wu:2021ibi}, which assumes a constant efficiency $f_\text{th}^\text{\knec} = 0.5$ for all channels.
Our improved \texttt{DT08} model addresses this issue to a large extent, but a residual uncertainty still persists at late times.
For $t\lesssim3$~days, the light curves are overestimated by no more than a factor of $40\%$.
This upper limit increases at later times.
The systematic uncertainty can in principle be reduced by introducing a time-dependent thermalization factor for the residual term, which must take into account the fission contributions and must decrease monotonically to suppress the deviations between \skynet\ and \ENDF.
Nonetheless, by these epochs the LTE assumption underlying our radiative-transport model becomes unreliable, and the thermalization details become a minor source of uncertainty.
This uncertainty does not affect the comparison carried out in the other sections of this paper, as the systematics are the same in all simulations.

\subsection{Effects of opacity models}
\label{subsec:res_opacity}

The (optical) opacity enters the luminosity term in Eq.~\eqref{eq:energy_cons}, and can influence the ejecta dynamics.
Consequently, a different opacity model could also indirectly affect the composition evolution.
However, since radiation contributes only a subdominant fraction of the energy balance (see Fig.~\ref{fig:check_cons}), a detailed multigroup treatment of the opacity is not necessary to accurately capture the hydrodynamics. 
The radiation term becomes relevant only at $t \gtrsim 1$~days, when the expansion is already fully homologous and changes in the composition of the outflow are driven only by nuclear decays.
An order-of-magnitude estimate of the gray opacity is therefore sufficient to correctly reproduce the ejecta dynamics and perform reliable nucleosynthesis calculations.
The relative difference between the abundances predicted by different opacity models always remain below a few~$0.1\%$, and the final mass fraction of lanthanides is essentially unchanged (see Table~\ref{tab:knecnn_runs}).

\begin{figure*}
	\centering 
	\includegraphics[width=\textwidth]{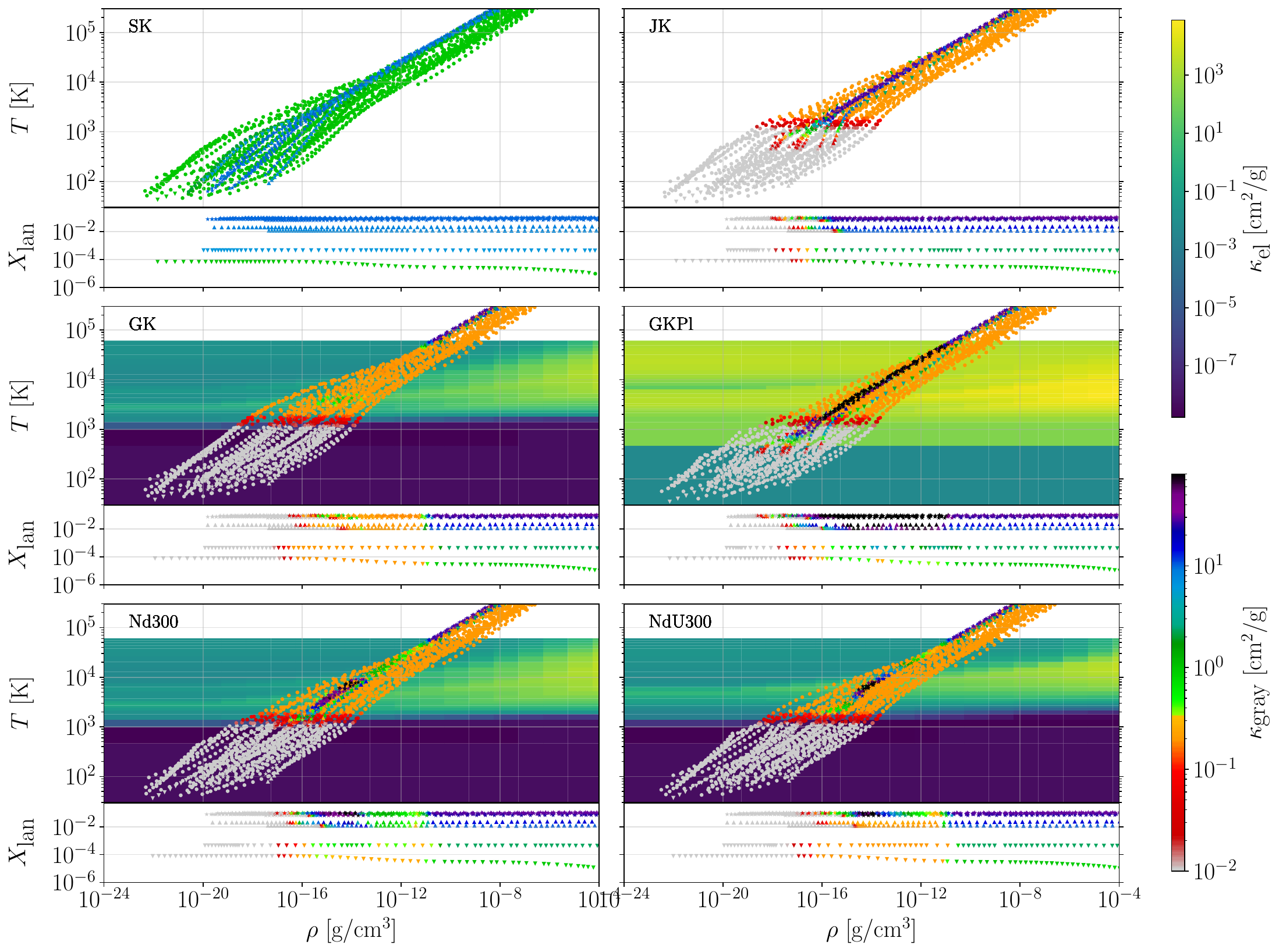}
	\caption{Evolution of the gray opacity $\kappa_\textrm{gray}$ for a sampling of fluid elements across all angles and depths, from simulations using different opacity models.
		\textit{Top row}: Runs using the analytical expression from \cite{Wu:2021ibi} (\textit{left}) or the prescription from \cite{Just:2021vzy} (\textit{right}).
		\textit{Central row}: Simulations with Rosseland (\textit{left}) or Planck (\textit{right}) gray opacities computed from the LANL data for neodymium as representative of the lanthanides.
		\textit{Bottom row}: Rosseland mean opacity from the multigroup models using only neodymium to represent lanthanides (\textit{left}) or also uranium as representative of actinides (\textit{right}).	
		The color bar is cut at $\kappa_\textrm{gray} = 10^{-2}$~cm$^2$g$^{-1}$ for easier visualization.
		For each simulation, the top and bottom subpanels show the trajectories of the fluid elements in the density-temperature and density-cumulative lanthanides mass fraction ($X_\text{lan}$) planes, respectively.
		Markers indicate different orders of magnitudes of $X_\text{lan}$, with circles for $X_\text{lan} < 5\times10^{-5}$, downward triangles for $10^{-5} < X_\text{lan} < 10^{-3}$, upward triangles for $10^{-3} < X_\text{lan} < 5\times10^{-2}$, and stars for $X_\text{lan} > 5\times10^{-2}$.
		For the simulations using only neodymium data, the background shows the single-element mean opacity $\kappa_{\textrm{el}}$ calculated from the LANL tables (Rosseland for \texttt{GK} and \texttt{Nd300}, Planck for \texttt{GKPl}).
		Similarly, the uranium Rosseland mean opacity is shown in the \texttt{NdU300} panel.}
	\label{fig:kappa_compare}
\end{figure*}

Figure~\ref{fig:kappa_compare} shows the evolution of the effective gray opacity for a representative set of fluid elements spanning the full \BLh\ ejecta for the opacity models introduced in Sec.~\ref{subsec:opacity}.
The W[22] model (used for the \texttt{SK} simulation) assigns to each fluid element a constant opacity determined only by its initial electron fraction.
At early times, corresponding to temperatures $T \gtrsim 2\times10^3$~K, this prescription overestimates the effective gray opacity in fluid elements that are lanthanide-poor and underestimates it in lanthanide-rich regions when compared to the multi–group results.
At late times, W[22] always overestimates the optical opacity.

The same reasoning applies when comparing the multigroup calculations with the J[22] model, employed for the \texttt{JK} run.
Although this prescription adapts better to the multigroup results based on neodymium (and uranium) data (simulations \texttt{Nd300} and \texttt{NdU300}), it still overestimates the opacity in the (equatorial) lanthanide-active ($X_\text{lan} \gtrsim 10^{-5}$) regions at temperatures around $T\sim10^3$~K.

By construction, the Rosseland (\texttt{LANL-R}) and Planck (\texttt{LANL-P}) mean opacities (simulations \texttt{GK} and \texttt{GKPl}) tend to respectively underestimate and overestimate the effective opacity of the medium when compared against the frequency-dependent (\texttt{LANL}) model.
As expected, the effective gray opacities of the lanthanide-producing fluid elements from the \texttt{GK} and \texttt{GKPl} simulations follow the structure of the underlying neodymium data. The trajectories with high initial electron fraction do not synthesize lanthanides, and their opacities are therefore set by the time-dependent opacity floor, consistent with the prescription of \cite{Just:2021vzy}.
The effective opacity of the frequency-dependent \texttt{Nd300} and \texttt{NdU300} simulations have a qualitatively different behavior with respect to the gray \texttt{GK} and \texttt{GKPl} calculations, showing a strongly enhanced peak around $T\sim5\times10^3$~K, corresponding to $0.1 \lesssim t~[\mathrm{days}] \lesssim 1$ depending on the fluid element.

The inclusion of uranium data (simulation \texttt{NdU300}) reduces the effective gray opacity of the $r$-process fluid elements, as the Rosseland opacity of the actinide outside the resonant region lies below our floor value of $\kappa_\mathrm{oth} = 0.2~\kappa_T~\mathrm{cm}^2\mathrm{g}^{-1}$ (used for the actinides in \texttt{Nd300}, where uranium is not included). In the \texttt{NdU300} run, the opacity peak at $T \sim 5 \times 10^3$~K is more localized, with a more gradual rise and an earlier decline, with respect to the \texttt{Nd300} simulation.

\begin{figure}
	\centering 
	\hspace*{-12pt}\includegraphics[width=0.5\textwidth]{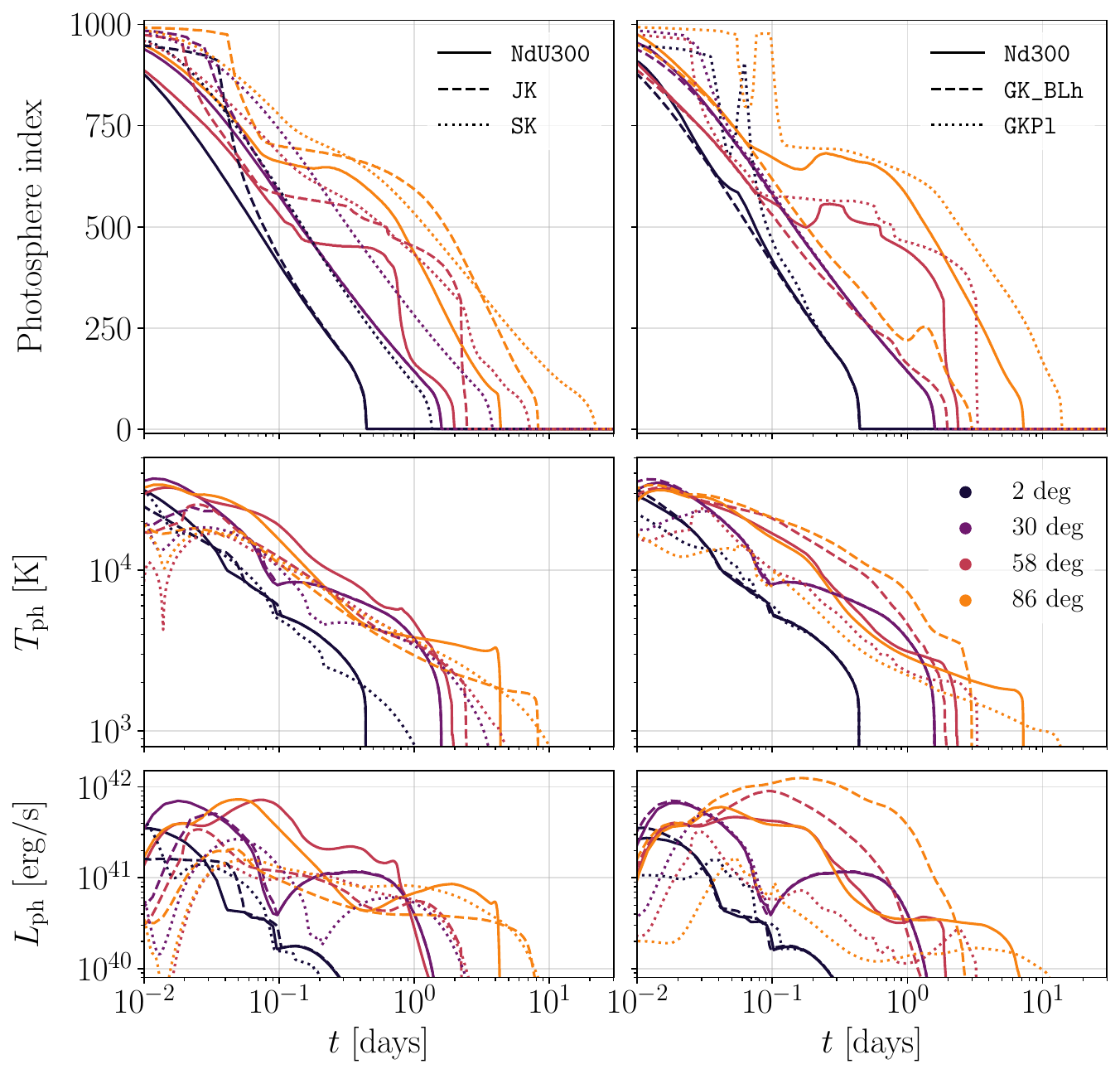}
	\caption{Position (\textit{top}), temperature (\textit{middle}) and spherically-symmetry-equivalent luminosity (\textit{bottom}) of the effective gray photosphere for the \BLh\ profile. Line styles and colors distinguish between different opacity models and angular sections, respectively.}
	\label{fig:photosphere}
\end{figure}

The opacity model shapes the evolution of the photosphere.
Figure~\ref{fig:photosphere} shows the effective gray photosphere for selected angular sections as a function of time.
The W[22] model always places the photosphere farther out in the ejecta than the \texttt{LANL} models (simulations \texttt{Nd300} and \texttt{NdU300}), resulting in a systematically colder photosphere.
At early times ($T \gtrsim 2\times10^3$~K), this occurs because W[22] overestimates the opacity in the lanthanide-poor regions, where the photosphere is initially located.
At later times, as the photosphere recedes into the $r$-process–active, lanthanide-rich (equatorial) layers, this model continues to overestimate the opacity across all initial electron fractions.

The J[22] prescription reproduces the early opacity behavior more accurately, though it still fails to capture the opacity drop around $T \sim 2\times10^3$~K.
Consequently, the differences in the photosphere evolution between this model and \texttt{NdU300} are more pronounced in the equatorial regions (which contribute the most to the EM emission) than in the polar ones.

The nonanalytical models exhibit non-monotonic opacity evolution in the lanthanide-rich elements.
This behavior arises either from the opacity peak around $T\sim5\times10^3$~K in the \texttt{LANL-R} and \texttt{LANL} cases, or from the activation of the atomic data at $T\simeq6\times10^4$~K in the \texttt{LANL-P} model (see Fig.~\ref{fig:kappa_compare}).
As a result, the photosphere is observed to bounce toward the outer regions of the ejecta at $t\sim1$~hour in the \texttt{GKPl} simulation, and later at $t\sim5$~hours in the \texttt{GK}, \texttt{Nd300}, and \texttt{NdU300} runs.
The photosphere location predicted by the multigroup calculations always lies between those obtained with \texttt{LANL-R} and \texttt{LANL-P}, which are known to respectively under- and overestimate the effective gray opacity.

\begin{figure*}
	\centering 
	\includegraphics[width=\textwidth]{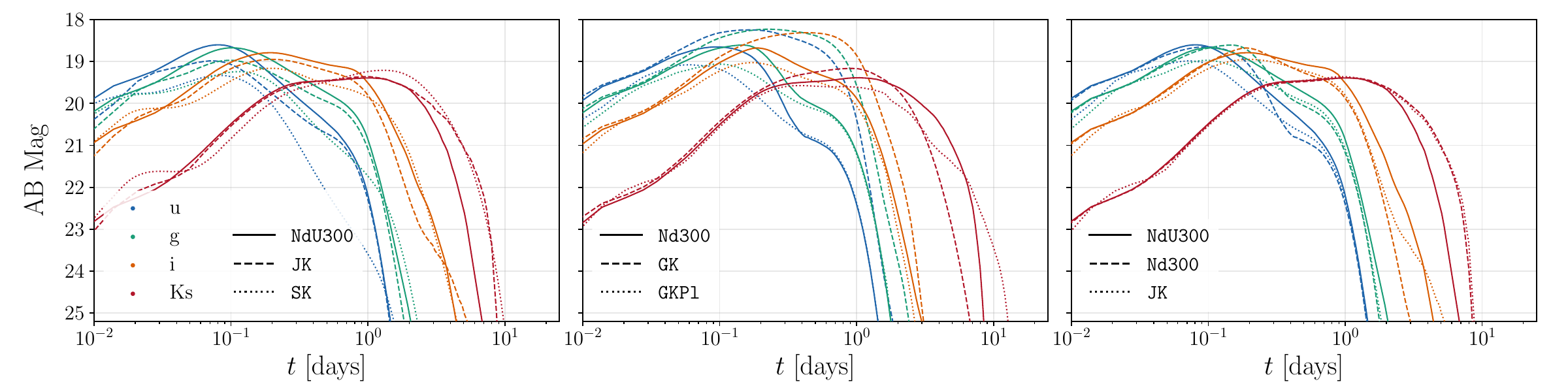}
	\caption{Predicted AB apparent magnitudes in the Gemini $u$, $g$, $i$ and $K_s$ bands for an observer at a polar angle of $30$~degrees and a distance of 40~Mpc.
	Comparison between a series of opacity models applied to the \texttt{BLh} profile.
	\textit{Left:} Frequency- and composition-dependent opacities based on atomic calculations for neodymium and uranium (solid lines), versus the \cite{Just:2021vzy} prescription (dashed) and the simple \cite{Wu:2021ibi} opacity (dotted).
	\textit{Center:} Opacities based on the neodymium atomic calculations, multigroup (solid lines) versus gray models (Rosseland or Planck, dashed and dotted).
	\textit{Right:} Frequency-dependent LANL opacities including uranium (solid lines) or only neodymium (dashed) compared against the analytical \cite{Just:2021vzy} model (dotted).}
	\label{fig:light_curves_OP}
\end{figure*}

In Fig.~\ref{fig:light_curves_OP}, we examine the impact of the opacity model on the kilonova light curves. The colder photosphere predicted by W[22] with respect to the \texttt{LANL} model results in a suppressed emission at $t \lesssim 1$~hour (in agreement with the bolometric luminosities reported in Table~\ref{tab:knecnn_runs}), with a stronger suppression at higher frequencies that makes the early-time signal appear redder.
Nevertheless, the photospheric luminosity and the optical-band magnitudes remain broadly comparable across models, as the larger photospheric radius partially compensates for the lower emissivity of the colder photosphere.
At later times, the artificially high constant gray opacity causes the ejecta to become completely optically thin only at comparatively late stages.
Therefore, the photospheric contribution to the luminosity persists for longer, thus delaying the decay of the infrared emission at $t\gtrsim1$~day.

As discussed above, J[22] performs better than W[22], but still overestimates the opacity of the lanthanide-rich ejecta.
As a result, the light curves from J[22] more closely resemble those produced with the \texttt{LANL} models than those from W[22], but they still exhibit a lower peak bolometric luminosity and a delayed shut-off of the photospheric emission and thus a more prolonged signal.

The central panel of Fig.~\ref{fig:light_curves_OP} compares simulations using gray (\texttt{GK} and \texttt{GKPl}) and multigroup (\texttt{Nd300}) opacities based on neodymium atomic calculations.
As expected, the Rosseland mean opacity performs better in the light-curve calculations at early times, when the ejecta is still optically thick.
At later times, the Planck opacity provides a better approximation for the kilonova emission, especially in the higher-frequency bands.
The two gray approximations bracket the actual effective opacity, and therefore generally bracket the multigroup light curves across all filters over most of the evolution.
In particular, the Planck mean misses the early blue/UV peak at $t\sim0.1$~days, whereas the Rosseland opacity predicts too bright and delayed magnitude peaks, especially in the higher-frequency filters.
The peak bolometric luminosity is consequently enhanced and delayed by the \texttt{LANL-R} model (see Table~\ref{tab:knecnn_runs}).
At late times ($t \gtrsim 10$~days), the Planck opacity performs worse again because of the delayed recession of the photosphere in a more opaque material (see Fig.~\ref{fig:photosphere}).
This effect dominates over the direct suppression of the emergent flux (\eg\ Eqs.~\eqref{eq:flux_eq}~and~\eqref{eq:lum_eq_ross}), leading to an inversion of the hierarchy of the models in the red filter at $t \gtrsim 5$~days.

The right panel of Fig.~\ref{fig:light_curves_OP} shows that J[22] reproduces multigroup results based on neodymium data more accurately than our gray, atomic-physics-based models.
This result highlights the necessity of employing frequency-dependent transport when using detailed atomic opacities, as compressing the information from line-resolved calculations into effective gray averages can be misleading to the point of being outperformed by analytical fitting models.

The original work of \cite{Kasen:2017sxr} employed atomic calculations only up to the lanthanides, which at least partially explains the disagreement with our most sophisticated opacity model that also includes uranium (\texttt{NdU300}).
\cite{Fontes:2019tlk} demonstrated differences between semi-relativistic and fully relativistic neodymium opacities.
Additional improvements in opacity modeling may be achieved by replacing theoretically calculated energy levels with NIST-calibrated values.
Incorporating these calibrated energies may affects the corresponding light curves and spectra at late times. %

When only neodymium is incorporated, the effective opacities predicted by the frequency-dependent simulations lie between the Rosseland and Planck mean values, but are qualitatively closer to the latter (see Fig.~\ref{fig:kappa_compare}).
Correspondingly, the J[22] approximation produces light curves more similar to \texttt{LANL-P}, particularly at late times, when the ejecta becomes optically thin.
However, the analytical prescription fails to capture the drop in effective opacity for the $r$-process fluid elements at $T\sim2-3\times10^4$~K.
As a result, it misses the luminosity peak at $t\sim1-3$~hours, caused by regions of the outflow that are hotter than their surroundings due to early ($t\sim1$~s) decays of freshly synthesized $r$-process elements \citep{Bernuzzi:2024mfx}.
This missing peak, already visible in the original publication by \cite{Just:2021vzy}, is more relevant for trajectories with higher lanthanide fractions.

Including uranium (\texttt{NdU300}) gives a brighter, earlier and, bluer kilonova peak (see also Table~\ref{tab:knecnn_runs}), a smoother decay in the blue/UV filters ($u$ and $g$), a prolonged emission in the red ($i$) filter, and a faster drop in the infrared ($K_s$), driven again by the more rapid recession of the photosphere.

\subsection{Impact of a polar jet}
\label{subsec:res_jet}

As a representative example of the effect that a jet as described in Sec.~\ref{subsec:GRB_model} can have on our results, we run again the \texttt{GK} model including a polar, structured jet.
We fix $\theta_j = 15$~degrees for the opening angle, within the range also explored in \cite{Hamidani:2019qyx} for the GRB associated with GW170817.
We start the energy injection at $t_\text{jet} = 200$~ms and choose for the jet the same duration as the observed GRB from \cite{LIGOScientific:2017zic}, $\Delta t_j = 100$~ms.
We impose for the isotropic energy parameter $E_0 = 10^{51}$~erg, compatible with the isotropic luminosity range given by \cite{Hamidani:2019qyx}. %

\begin{figure}
	\centering 
	\hspace*{-12pt}\includegraphics[width=0.5\textwidth]{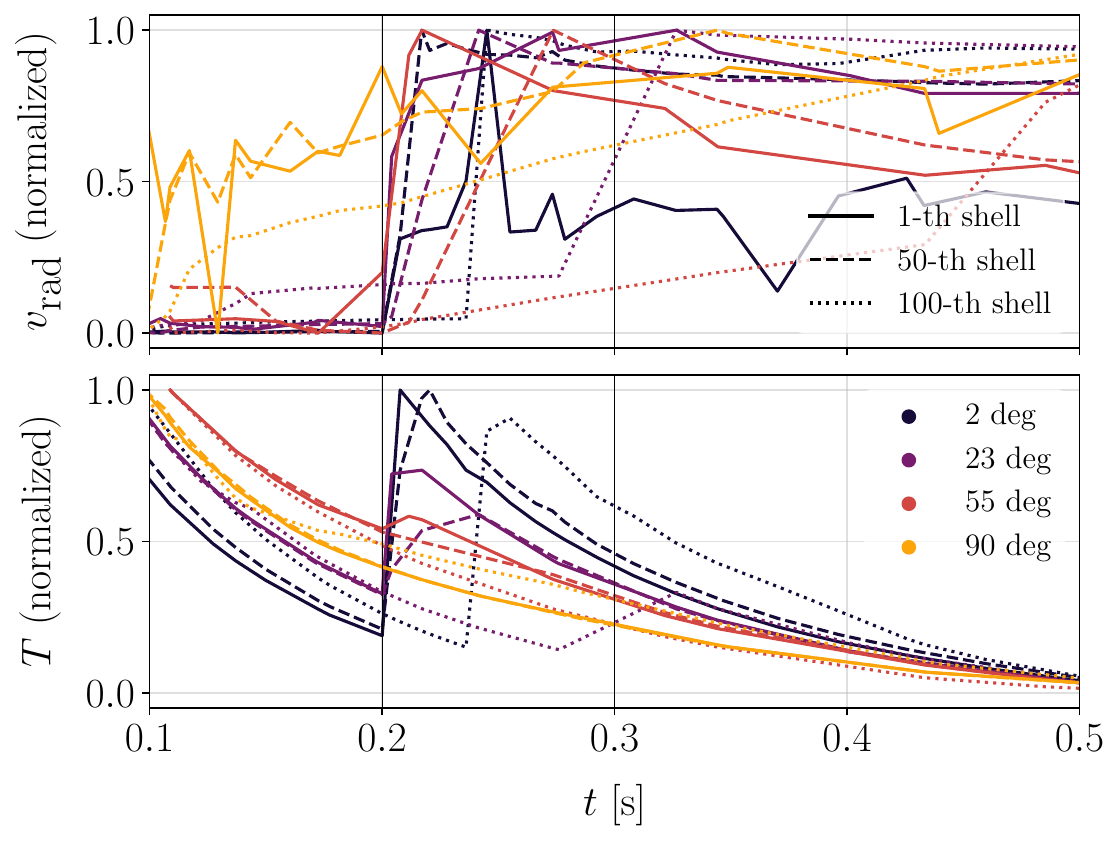}
	\caption{Radial velocity (\textit{top}) and temperature (\textit{bottom}) for a set of fluid elements from the \texttt{GK} simulation including the polar jet.
	Both quantities are rescaled to the interval $[0,1]$ for visualization purposes.
	Colors and styles correspond to different angular sections or mass-shell indices, respectively. The vertical black lines indicate the duration of the jet.}
	\label{fig:jet_shock}
\end{figure}

In Fig.~\ref{fig:jet_shock}, we show the evolution of the radial velocity and temperature, each rescaled to the interval $[0,1]$ over the plotted time range, for a few representative fluid elements on the timescales relevant for the GRB. The activation of the thermal bomb generates a shock visible in the radial velocity of the innermost mass shells.
In the most polar regions, the same fluid elements also experience a rapid increase in temperature.
The impact of the jet becomes progressively weaker at larger polar angles, where the deposited energy is smaller.
For outer mass shells, the response is delayed, reflecting the finite propagation time of the shock through the ejecta.

\begin{figure}
	\centering
	\includegraphics[width=0.49\textwidth]{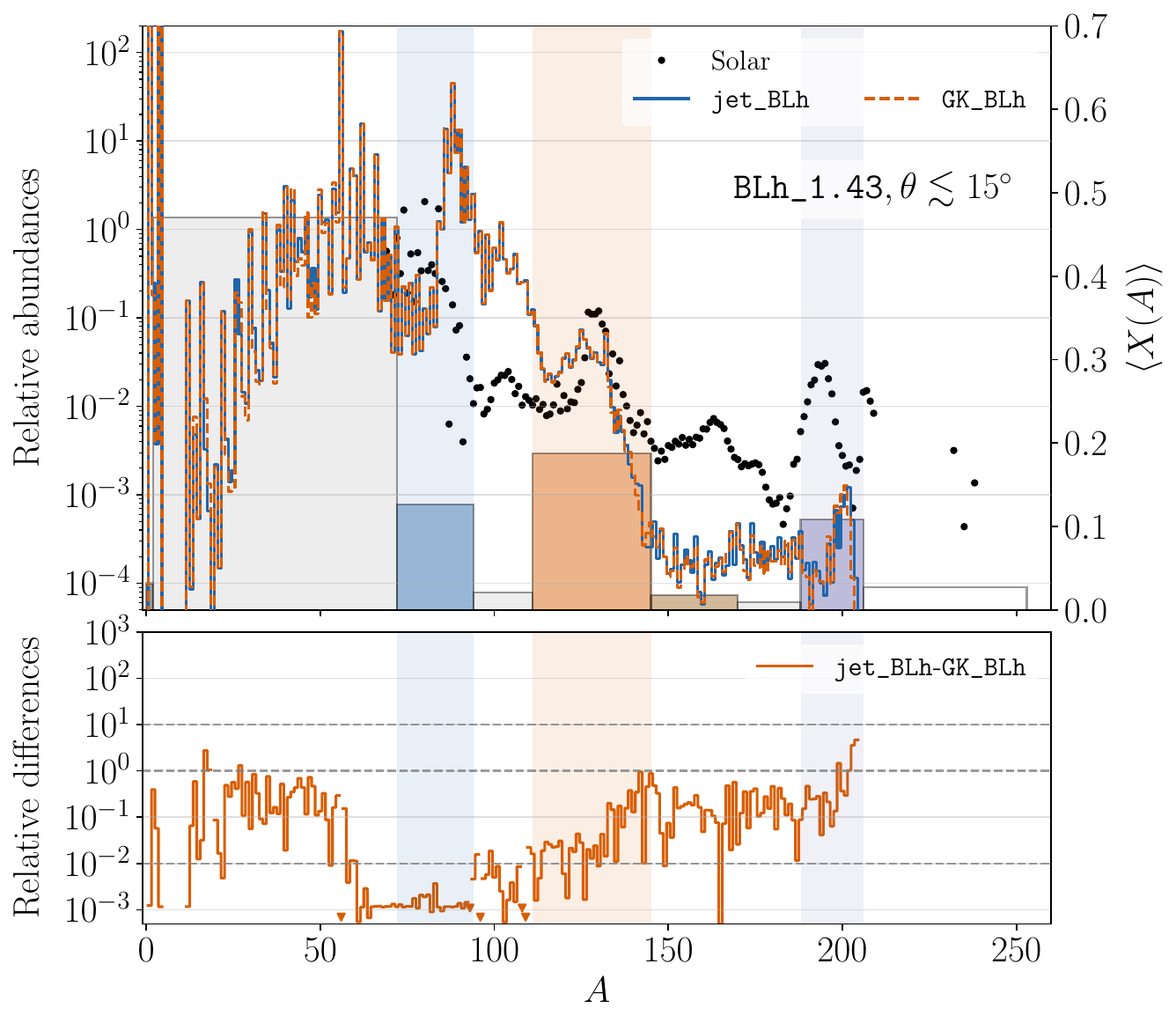}
	\caption{Same as Fig.~\ref{fig:yields_onlineNN}, but now comparing the \texttt{GK} simulation against the same run but including a polar jet (\texttt{jet}), and considering only the material ejected within $\theta \lesssim 15$~degrees.
	The abundances are normalized such that the cumulative abundance of the second $r$-process peak ($111 \leq A \leq 145$) is unity.
	The histogram shows the cumulative mass fractions for the simulation with the polar jet.}
	\label{fig:yields_jet_polar}
\end{figure}

As already commented in \cite{Magistrelli:2024zmk}, the impact of the jet on the globally averaged nucleosynthesis is negligible.
Figure~\ref{fig:yields_jet_polar} shows the nuclear yields from the polar regions ($\theta \lesssim 15$~degrees) of the \BLh\ ejecta in the structured-jet and no-jet configurations.
Because this angular cut excludes the neutron-rich equatorial outflow, the resulting nucleosynthesis is dominated by the production of elements up to the first peak.
The nuclear yields are essentially unchanged by the introduction of the jet, with typical discrepancies on the order of a few tens of percent or less. %

\begin{figure}
	\centering
	\includegraphics[width=0.49\textwidth]{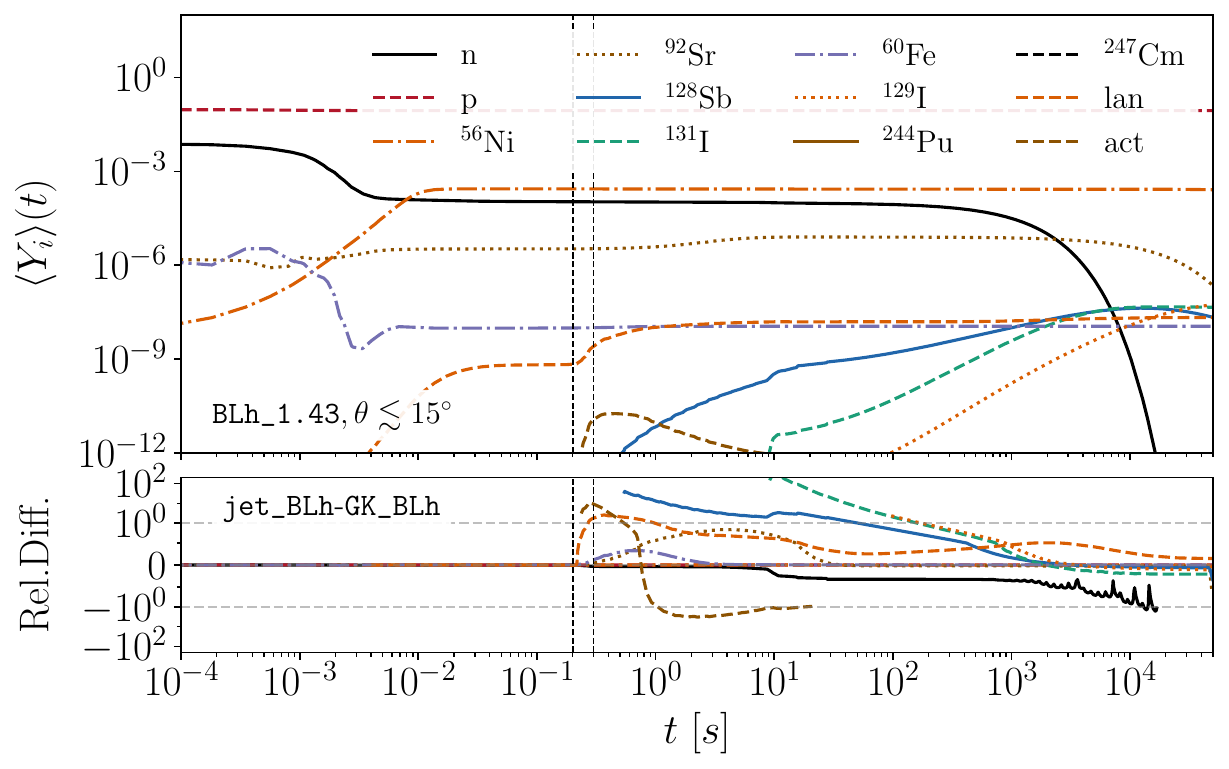}
	\caption{Time evolution of the mass-weighted abundances for selected isotopes and cumulative lanthanides and actinides abundances in the \texttt{GK} simulation, considering only matter ejected at latitudes $\theta \lesssim 15$~degrees.
	The vertical black lines mark the time interval during which the thermal bomb is active.
	\textit{Top panel}: Absolute abundances from the simulation including a polar jet (\texttt{jet}).
	\textit{Bottom panel}: Relative differences with respect to the same simulation, but without polar jet. 	The horizontal dashed lines indicate factors-of-unity deviations and the case $\langle Y_i \rangle = \langle Y_i^{pp} \rangle$.
	Differences are shown only at times when~$\langle Y_i \rangle > 10^{-12}$.}
	\label{fig:spec_abs_jet_polar}
\end{figure}

Figure~\ref{fig:spec_abs_jet_polar} displays time evolution of selected isotopes, lanthanides and actinides.
The overall behavior is characteristic of an only mildly neutron-rich environment, with no significant production of lanthanides, negligible actinides, and only minor synthesis of elements heavier than $A\gtrsim90$.
The evolution is similar irrespectively of the jet.
Minor differences include a boost in the lanthanides production during the jet activity, and an earlier appearance of $^{128}$Sb and $^{131}$I on longer timescales.
The production of the small amount of actinides is faster in the jet case, while their abundance drops at a comparable rate in the both simulations.

The energy deposited by the thermal bomb in our simulation is not enough to significantly affect the kilonova light curves.
Based on the small effects observed in our test, we expect that a more efficient energy deposition or a more powerful jet would enhance the early ($t \sim$~hours), high-frequency optical emission, while producing a dimmer kilonova at later times ($t \gtrsim 1$~day), consistent with the findings of \cite{Nativi:2020moj, Magistrelli:2024zmk}.

\subsection{Best model comparison}
\label{subsec:res_best_comparison}

In this section, we summarize the combined effects of all the improvements that we implemented on top of the code from \cite{Wu:2021ibi}.
We compare our most advanced setup (\texttt{NdU300}), featuring the in-situ NN, detailed thermalization, multigroup radiative transfer, and atomic-physics based opacities, against the results obtained using the 2D ray-by-ray extension of original version of the code (\texttt{Apr2\_BLh}).
For the latter, we also implement the generalized fits for the nuclear power discussed in Sec.~\ref{subsec:nucl_pow_fits}.

As already mentioned in the previous sections, the improved thermalization and opacity prescriptions do not significantly affect the composition evolution of the ejecta.
Slightly larger discrepancies appear in the nucleosynthesis yields and abundances evolution with respect to those described in Sec.~\ref{subsec:res_onlineNN}, but the physical considerations remain the same.
The introduction of the in-situ NN is the dominant source of variation in the nucleosynthesis results, and a detailed description of the hydrodynamic evolution of the ejecta is essential for reliable nucleosynthesis predictions.
In particular, assuming homologous expansion is not safe on $r$-process timescales, leading to deviations greater than unity for most mass numbers and a shifted third $r$-process peaks (see the green and red curves in the bottom panels of Fig.~\ref{fig:yields_onlineNN}).
On the other hand, calculating the nuclear power on the fly and including its back-reaction on the hydrodynamics has a comparatively minor impact, typically at the $\lesssim 20\%$ level (see the orange lines in the same panels).
For the purpose of studying element formation, the hydrodynamics of the ejecta can therefore be reproduced well enough by performing simulations with accurately calibrated nuclear power fits.
A post-processing NN can then recover the results of the online calculations within a few $10\%$, provided that the set of tracers is sufficiently large to properly resolve both the dynamics and the initial composition distribution of the ejecta.

\begin{figure}
	\centering 
	\hspace*{-10pt}\includegraphics[width=0.5\textwidth]{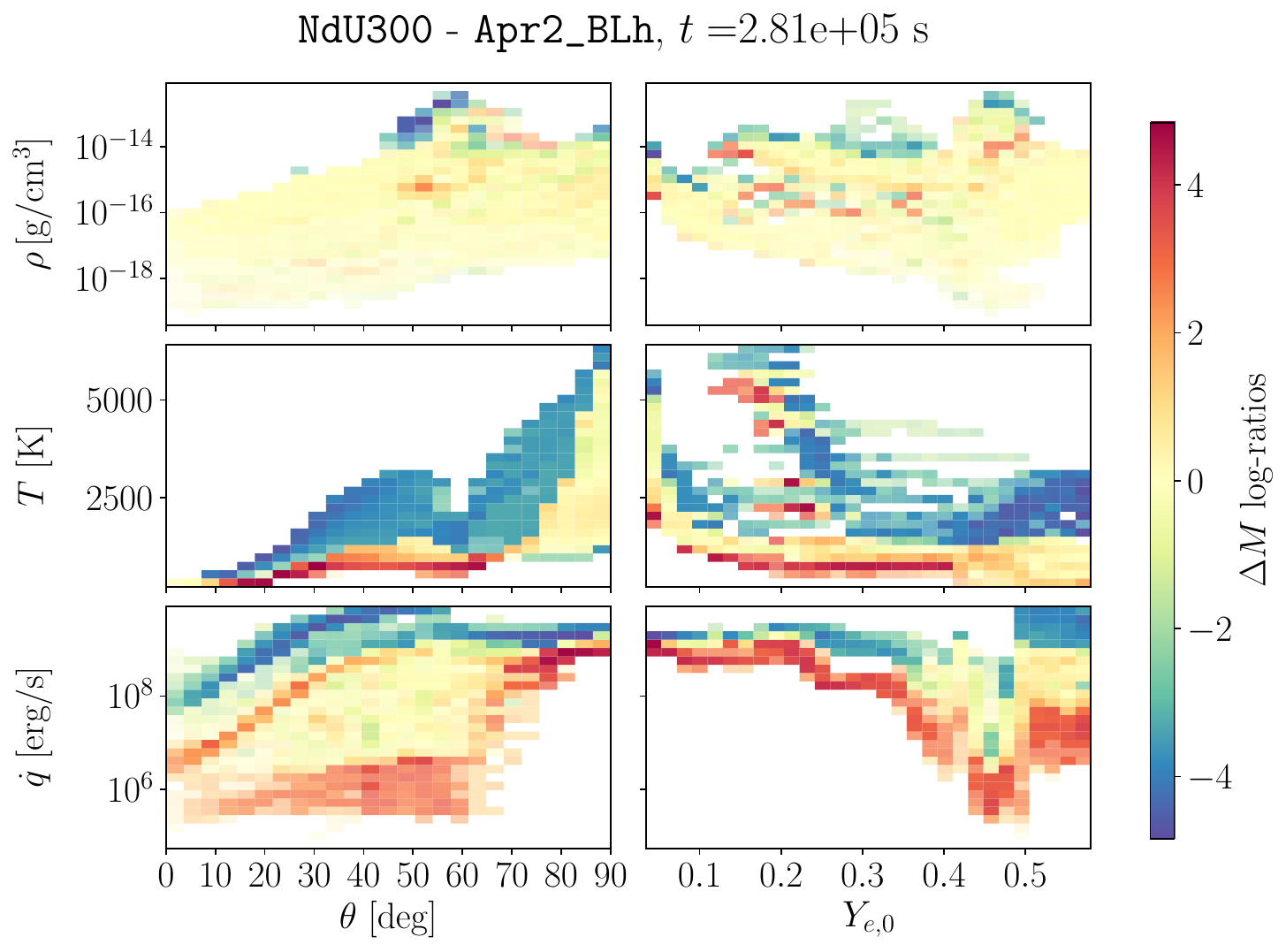}
	\caption{Late-time ($t\simeq3$~days) ratios between the mass-weighted histograms of the density, temperature and instantaneous thermalized heating rate obtained from the \BLh\ runs using the in-situ NN and our most advanced thermalization and opacity prescriptions (\texttt{NdU300}), or from the 2D ray-by-ray extension of the configuration of \cite{Wu:2021ibi} (\texttt{Apr2\_BLh}).
		The left [right] column shows the ratios as functions of the polar angle [initial electron fraction].
		Bins containing less than $10^{-5} M_\odot$ in both simulations are plotted with increasing transparency for decreasing mass content (considering the run with most mass in that bin).
		Thus, the color encodes the value of the ratio in $\Delta M$, while the saturation reflects the highest between the two original values $\Delta M$ (cf. Fig.~\ref{fig:hydro_onlineNN})
	}
	\label{fig:hydro_bestcomp}
\end{figure}

The \texttt{LANL} opacity model has a negligible impact on the hydrodynamics of the ejecta, as already discussed in Sec.~\ref{subsec:res_opacity}.
However, once the composition-dependent thermalization made available by the in-situ NN is activated, its combination with the online nuclear calculations leads to qualitative differences in the thermodynamic evolution of the fluid elements.
Figure~\ref{fig:hydro_bestcomp} shows the ratios between the two simulations for the histograms of density, temperature and thermalized heating rate at $t\sim3$~days, plotted as functions of polar angle and initial electron fraction.
The deviations in the density profile develop during the transition to the homologous expanding phase, at times $0.1 \lesssim t~[\mathrm{s}] \lesssim 1$.
These differences are slightly amplified, but remain qualitatively comparable to those caused by enabling the in-situ NN alone (see Fig.~\ref{fig:hydro_onlineNN}), since the average thermalization efficiency is of the same order of the simple $f_\text{th}^\text{\knec}$ at early times.
Once homologous expansion is established, the deviations in the density profiles freeze and remain fixed until the end of the simulation.

The impact of the in-situ NN and the \texttt{DT08} thermalization scheme is more apparent in the temperature evolution.
The two temperature profiles begin to differ after a few hours, and their ratios are initially comparable to those arising solely from the online nuclear calculations.
However, at late times ($t\gtrsim$~day), the thermalization efficiency predicted by the advanced model becomes orders of magnitude smaller than the fixed $f_\text{th}^\text{\knec}$.
Although the nuclear-power fits generally reproduce the correct order of magnitude of the released energy, the simpler thermalization scheme significantly overestimates the heating rate at late times.
As a result, it yields temperatures that are a factor of a few higher than those obtained with the composition-dependent thermalization prescription.

\begin{figure}
	\centering 
	\hspace*{-10pt}\includegraphics[width=0.5\textwidth]{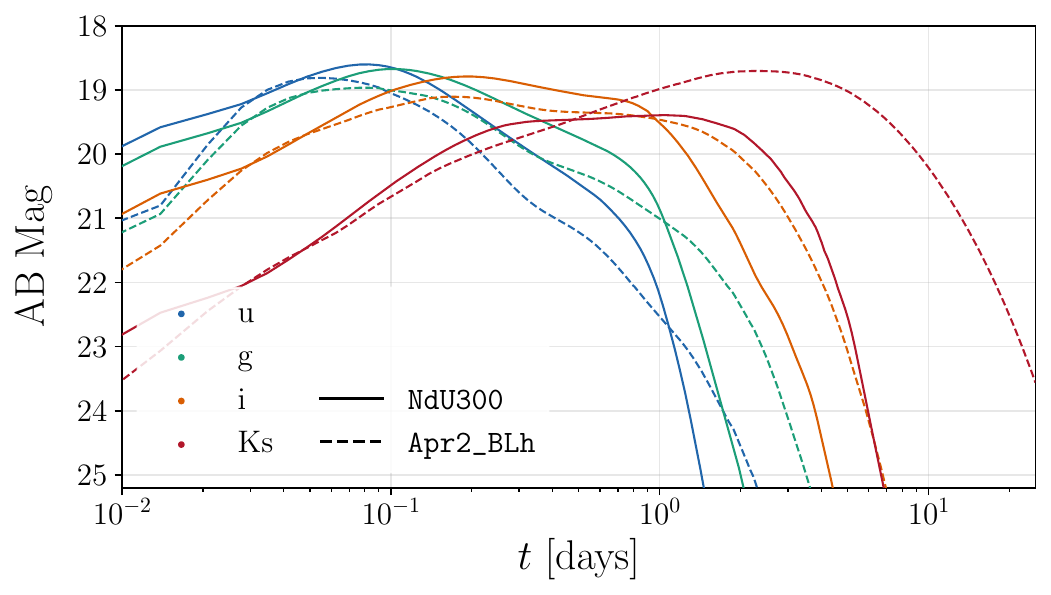}
	\caption{Predicted AB apparent magnitudes in the Gemini $u$, $g$, $i$ and $K_s$ bands for an observer at a polar angle of $30$~degrees and a distance of 40~Mpc.
	Comparison between the \BLh\ simulations using in-situ NN and our most advanced models for thermalization and opacity (solid lines), or the 2D ray-by-ray extension of the setup of \cite{Wu:2021ibi}, with their constant prescription for thermalization and opacity, and our updated nuclear power fits (dashed lines).}
	\label{fig:light_curves_bestcomp}
\end{figure}

The differences in temperature evolution and thermalized heating rates have an immediate impact on the predicted light curves.
Figure~\ref{fig:light_curves_bestcomp} displays the kilonovae associated with the \texttt{Apr2\_BLh} and \texttt{NdU300} simulations, both showing the characteristic blue/UV peak at $t\sim1$~hour and the later red emission.
At early times ($t\lesssim0.1$~day), the average thermalization efficiency is comparable between the two simulations, and higher in the \texttt{DT08} scheme.
At the same time, the multigroup opacities drive a faster recession of the photosphere.
Both effects contribute to a hotter photosphere than in the simpler model.
The early kilonova, which is dominated by the photospheric emission, is therefore brighter and peaks at higher frequencies.
Moreover, the more efficient thermalization keeps the ejecta hotter for longer, shifting the peak at later times, around $t\sim2$~hours.
The thermalization efficiency drops exponentially for $t\gtrsim0.1$~day.
At late times ($t\gtrsim1$~day), the advanced model predicts a smaller contribution to the light curves from the energy thermalized in the optically thin layers of the outflow.
The material is overall colder, and so is the photosphere (despite its faster recession due to the multigroup opacities).
Both the photospheric emission and the contribution to the luminosity from the optically thin layers are therefore suppressed by the composition-dependent thermalization, explaining the steeper late-time decline.
The \texttt{LANL} opacity model further reinforces this suppression because of the faster recession of the photosphere, which leads to an earlier drop in the infrared bands.

We stress again that the use of our most advanced models for thermalization and opacity relies on the nuclear information provided by the in-situ NN.
Our results can be employed to construct effective fits for the heating rate or analytical opacity prescriptions, following the approach of \eg\ \cite{Just:2021vzy}.
However, because the thermodynamics of the ejecta strongly depends on the properties of the original binary system, a posteriori consistency checks are recommended when possible (\eg\ on a selected region of the outflow).

\section{Conclusions}
\label{sec:conclusion}

In this paper, we carried out a systematic analysis of the impact of fully coupling nuclear calculations to the hydrodynamic evolution of ejecta from BNSMs on their dynamics, nucleosynthesis, and kilonova light curves.
We evolved ejecta profiles extracted from ab-initio NR simulations up to $t\sim30$~days, and explored different nuclear burning, thermalization, and opacity models.
Our most advanced setup couples the information provided by the in-situ NN to composition-dependent, nuclear-informed thermalization calculations and atomic-physics based, frequency-dependent opacities.

By fully coupling the hydrodynamics and nuclear evolution, we showed that prescribing a simplified exponential-plus-homologous expansion for the ejecta at early times ($t\lesssim500$~ms) leads to inaccurate predictions of the composition evolution and final yields.
However, the back-reaction of the nuclear power calculated on the fly only gives minor corrections to the density evolution, as the radial explosion is dominated by the kinetic energy of the outflow. Simulations in higher dimensions would be needed to asses whether more accurate nuclear-power calculations could be relevant for the motion along the polar and azimuthal directions.

The in-situ NN qualitatively impacts the early-time ($t\sim1$~hour) kilonova light curves by altering the location and temperature of the photosphere.
As a result, the emission is dimmer, redder, and peaks later compared to models using analytical fits for the nuclear power.
These simpler prescriptions perform relatively better in approximating the light curves at later times.
However, employing an in-situ NN also enables the use of composition-dependent thermalization and opacity models, whose significant qualitative effects on the EM emission are summarized in the remainder of this section.
The global corrections introduced by the online nuclear calculations to the density evolution of the outflow have negligible dynamical effects.

Composition-informed thermalization efficiencies are essential for accurately modeling the kilonova brightness and color evolution.
Using a constant thermalization factor typically predicts a too dim and red kilonova at $t\sim$~hours, and strongly overestimates the late-time EM emission from the optically thin regions of the outflow.
The thermalization treatment does not significantly affect the nucleosynthesis predictions.
A constant thermalization factor only gives $\sim 10\%$ deviations on the final abundance distribution.

The opacity model also qualitatively affects the kilonova emission.
Analytical prescriptions fail to capture the non-monotonic evolution of the effective opacities and typically overestimate them in the regions that determine the position of the photosphere.
As a result, they predict a more extended (and thus colder) photosphere.
The kilonova is therefore slightly dimmer and shifted toward redder frequencies at early times ($t\lesssim1$~hour), while the delayed recession of the photosphere prolongs the late-time red emission at $t\gtrsim5$~days.
The work from \cite{Just:2021vzy} showed that composition-dependent opacity fits can qualitatively reproduce the bolometric luminosity and the late-time red kilonova, but their analytical prescription overestimates the opacity at early times and is thus prone to miss early ($t\sim$~hours) blue peaks.
When actinides (represented here by uranium) are included in addition to lanthanides (represented by neodymium), the kilonova exhibits a brighter and bluer early peak ($t\sim$~hours), a smoother decay in the high-frequency bands, and a faster drop in the late-time red emission.
The opacity model has a negligible impact on the nucleosynthesis calculations, with discrepancies of $\lesssim 10^{-3}$ between different prescriptions.
Since the radiation term is subdominant in the energy equation, the hydrodynamic evolution of the ejecta is already well captured by gray, analytical approximations.

The effect of a polar jet injecting additional energy into the outflow is negligible for all global observables for a realistic set of parameters.
More energetic jets or a more efficient release of their energy into the ejecta are expected to produce a brighter and bluer kilonova at early times ($t\sim$~hours), followed by a faster decay on timescales of days.

The overall improvements enabled by the introduction of an in-situ NN can be summarized as follows.
Resolving the detailed hydrodynamical evolution of the outflow during the first several hundred milliseconds is essential for obtaining reliable nucleosynthesis predictions.
Calibrated analytical fits for the nuclear power, together with constant prescriptions for the thermalization efficiency and optical opacity, are sufficient to capture the bulk dynamics of the outflow.
However, a more accurate thermalization scheme is required to correctly reproduce its temperature evolution.
Moreover, we find that both the composition-dependent thermalization scheme and the multigroup opacity treatment are necessary to accurately model the kilonova emission. %

Due to the strong velocity gradients observed in the expanding ejecta, Doppler corrections are expected to modify the observed kilonova.
Such corrections also enter in the luminosity term in the energy equation.
However, due to the subdominant contribution of the luminosity term to the dynamics (see Fig.~\ref{fig:check_cons}), we expect only a minor impact on the dynamical evolution of the outflow.
In \knecnn, we implemented Doppler corrections in the luminosity term of the energy equation, in the photosphere tracking, and in the light-curve calculations.
These corrections are not included in the results presented in this paper.
Preliminary tests show a significant increase in execution time mainly in the polar regions, where localized shocks appear and slow down calculation.
As expected, our initial investigation indicates an almost negligible impact on the thermodynamic evolution of the outflow and on the nucleosynthesis.
The kilonova emission, however, is generally suppressed when Doppler effects are included.
The early ($t\sim$~hours) blue peak is damped, particularly at high frequencies, while the infrared emission is only mildly reduced, apart for short-lived luminosity bursts at $t\lesssim15$~hours.
We defer a more detailed analysis of the Doppler effects to a future publication.

In this paper, we used a Lagrangian approach and we did not investigate the role of composition mixing.
Some degree of mixing is physically expected, and its relevance is set by the Damk\"oler number $D_\alpha$, expressing the ratio between the mixing and the (nuclear) reaction timescales \citep{Dimotakis:2005}.
For $D_\alpha \sim 1$, composition mixing and nuclear reactions cannot be decoupled and might affect each other.
Lagrangian codes usually assume $D_\alpha \gg 1$ (mixing is not relevant), while Eulerian calculations typically overestimate the impact of mixing due to numerical dissipation.
In \cite{Zhai:2026uwn}, the effect of radial mixing is found to negligibly affect the global nucleosynthesis and light curves.
In the future, we plan to deepen this analysis using the more advanced opacity models introduced in this paper and to extend it to simulations in higher dimensions.

\section*{Acknowledgments}
We thank Leonardo Chiesa for making available for this work the updated version
of the \skynet\ grids used to define the nuclear power fits, and 
Ryan Wollaeger and Stephan Fritzsche for valuable discussions on atomic opacities.

FM acknowledges support from the Deutsche Forschungsgemeinschaft
(DFG) under Grant No. 406116891 within the Research Training Group
RTG 2522/1.
FM also thanks the Trento Institute for Fundamental Physics and Applications (TIFPA) for its hospitality during the development of the project.
SB acknowledges funding from the EU Horizon under ERC Consolidator Grant,
no. InspiReM-101043372 and from the Deutsche Forschungsgemeinschaft, DFG,
project MEMI number BE 6301/2-1.
AP is supported by the
European Union under NextGenerationEU, PRIN 2022
Project No. 2022KX2Z3B.
MJ acknowledges from the Deutsche Forschungsgemeinschaft, DFG,
project MEMI number BE 6301/2-1.

Simulations were performed on the ARA and DRACO clusters at Friedrich Schiller
University Jena, on the supercomputer SuperMUC-NG at the Leibniz-
Rechenzentrum (LRZ, \url{www.lrz.de}) Munich, and on the national HPE
Apollo Hawk at the High Performance Computing Center Stuttgart (HLRS).
The ARA cluster is funded in part by DFG grants INST 275/334-1
FUGG and INST 275/363-1 FUGG, and ERC Starting
Grant, grant agreement no. BinGraSp-714626. 
The authors acknowledge the Gauss Centre for Supercomputing
e.V. (\url{www.gauss-centre.eu}) for funding this project by providing
computing time on the GCS Supercomputer SuperMUC-NG at LRZ
(allocations {\tt pn36ge} and {\tt pn36jo}).
The authors acknowledge HLRS for funding this project by pro-
viding access to the supercomputer HPE Apollo Hawk
under the grant number INTRHYGUE/44215.

\bibliographystyle{aa}

\clearpage

\begin{appendix}

\section{Radiative transport equations}
\label{app:rad_transp_eq}

\newtxt{For matter in LTE (but not necessarily in equilibrium with radiation),} the transport equation for the specific intensity $I_\nu(\vec{r}, \vec{\Omega}, t)$ of radiation has the form \cite[\eg][]{Levermore:1981}
\begin{equation}
	\label{eq:rad_transp}
	\frac{1}{c} \frac{\partial I_\nu}{\partial t} + \vec{n}\cdot\nabla I_\nu + \sigma_\nu I_\nu =
	\frac{c}{4\pi} (\sigma_\nu^a \epsilon_\nu^P + \sigma_\nu^s \epsilon_\nu^r) \,,
\end{equation}
where $c$ is the speed of light, $ \vec{n}$ is the direction of propagation, \newtxt{$\sigma_\nu^{a,s} (\vec{r},t)$ are the absorption and scattering coefficients}, with $\sigma_\nu = \sigma_\nu^a + \sigma_\nu^s$ \newtxt{and $\sigma_\nu = \kappa_\nu \rho$, $\rho$ is the density, and $\kappa_\nu$ is the opacity}.
The \newtxt{specific energy density of the radiation field} is $\epsilon_\nu^r(\vec{r}, t) \equiv c^{-1} \int_{4\pi} d\Omega I_\nu$, with $\Omega$ solid angle, and $\epsilon_\nu^P = 4\pi B_\nu / c$ is the blackbody energy density. Integrating Eq.~\eqref{eq:rad_transp} over $d\Omega$ gives
\begin{equation}
	\label{eq:rad_transp_integr}
	\frac{\partial\epsilon_\nu^r}{\partial t} + \nabla \cdot \vec{F}_\nu =
	c \sigma_\nu^a (\epsilon_\nu^P - \epsilon_\nu^r) \,,
\end{equation}
where $\vec{F}_\nu(\vec{r}, t) \equiv \int_{4\pi} d\Omega \vec{n} I_\nu$.
If scattering is neglected, $\sigma_\nu^s = 0$, and Eq.~\eqref{eq:rad_transp} reduces to the Kirchhoff law for radiative transport\begin{equation}
	\label{eq:kirchhoff}
	\frac{1}{c} \frac{\partial I_\nu}{\partial t} + \vec{n}\cdot\nabla I_\nu =
	\kappa_\nu \rho (B_\nu - I_\nu) \,,
\end{equation}
Eq.~\eqref{eq:rad_transp} is exact and more general, but integrating Eq.~\eqref{eq:kirchhoff} over $d\Omega$ still leads to the same Eq.~\eqref{eq:rad_transp_integr}.

Assuming isotropic and quasi-stationary radiation (\ie\ radiation locally described by the time evolution of $T$ and $\rho$), one can manipulate Eq.~\eqref{eq:rad_transp_integr} to obtain the isotropic diffusion approximation (IDA) limit:\begin{align}
	\label{eq:flux_IDA}
	\vec{F}_\nu &= -\frac{c}{3 \sigma_\nu^a} \nabla \epsilon_\nu^r \,,\\
	\nabla \cdot \vec{F}_\nu &= c \sigma_\nu^a (\epsilon_\nu^P - \epsilon_\nu^r) \,.
\end{align}
Radiation-matter equilibrium is a sufficient (since it implies IDA) but not necessary condition for this approximation, for which only isotropic radiation on distances smaller than the mean free path $(\sigma_\nu^a)^{-1}$ of the radiation is required.
If \newtxt{radiation-matter equilibrium is assumed (single-temperature diffusion approximation)}, then $\epsilon_\nu^P = \epsilon_\nu^r$ and thus
\begin{equation}
	\label{eq:flux_LTE}
	\vec{F}_\nu = -\frac{c}{3 \kappa_\nu \rho} \nabla \epsilon_\nu^P
		= -\frac{4\pi}{3 \kappa_\nu \rho} \frac{\partial B_\nu}{\partial T} \nabla T \,.
\end{equation}

The treatment in \cite{Levermore:1981} is more general than IDA, as it develops the calculation starting from Eq.~\eqref{eq:rad_transp_integr} without assuming isotropic diffusion.
In particular, they introduce the normalized specific intensity $\psi_\nu$ such that $I_\nu = c \epsilon_\nu^r \psi_\nu$. In the general case, Eq.~\eqref{eq:rad_transp} becomes
\begin{equation}
	\label{eq:rad_transp_gen}
	\frac{1}{c} \frac{\partial (\epsilon_\nu^r \psi_\nu)}{\partial t} + \vec{n}\cdot\nabla (\epsilon_\nu^r \psi_\nu) + \sigma_\nu \epsilon_\nu^r \psi_\nu =
	\frac{1}{4\pi} (\sigma_\nu^a \epsilon_\nu^P + \sigma_\nu^s \epsilon_\nu^r) \,.
\end{equation}
Its angular integral still gives Eq.~\eqref{eq:rad_transp_integr}, as $\psi_\nu$ is normalized as $\int_{4\pi} d\Omega \psi_\nu = 1$.
Using only the assumption of a slowly varying normalized intensity, $\partial_t \psi_\nu + c \vec{n} \cdot \nabla \psi_\nu = 0$, one can find
\begin{equation}
	\psi_\nu = \frac{1}{4\pi} \frac{1}{R_\nu \coth R_\nu - \vec{n} \cdot \vec{R}_\nu} \,,
\end{equation}
where
\begin{equation}
	\vec{R}_\nu = - \frac{\nabla \epsilon_\nu^r}{\sigma_\nu \omega_\nu \epsilon_\nu^r} \,,
\end{equation}
$R_\nu = |\vec{R}_\nu|$, and
\begin{equation}
	\omega_\nu = \frac{\sigma_\nu^a \epsilon_\nu^P + \sigma_\nu^s \epsilon_\nu^r}{\sigma_\nu \epsilon_\nu^r}
\end{equation}
is the effective albedo.
The flux is given by
\begin{equation}
	\vec{F}_\nu = - \frac{\lambda_\nu c}{3 \sigma_\nu \omega_\nu} \nabla \epsilon_\nu^r \,,
\end{equation}
with $\lambda_\nu$ the flux limiting factor rationally approximated by Eq.~\eqref{eq:flux_limiter}.

In IDA, $R_\nu \to 0$, $\psi_\nu \to (4\pi)^{-1}$ and $\lambda_\nu \to 1$.
\newtxt{In the single-temperature diffusion approximation}, $\epsilon_\nu^r = \epsilon_\nu^P$, $\omega = 1$, and
\begin{equation}
	\label{eq:R_vector}
	\vec{R}_\nu = - \frac{\nabla B_\nu}{\kappa_\nu \rho B_\nu} \,.
\end{equation}
If the flux limiting factor is kept explicit, one gets
\begin{equation}
	\vec{F}_\nu = -\frac{\lambda_\nu c}{3 \sigma_\nu} \nabla \epsilon_\nu^P
	= -\frac{4\pi \lambda_\nu}{3 \kappa_\nu \rho} \frac{\partial B_\nu}{\partial T} \nabla T \,.
\end{equation}
In spherical symmetry (a single ray-by-ray section is an effective 1D problem), $\nabla T = dT/dr = 4\pi r^2 \rho\, dT/dm$, and thus the flux and flux limiter can be written as in Eq.~\eqref{eq:flux_eq}-\eqref{eq:R_factor}.

In the LTE gray radiative transport approximation of \cite{Bersten:2011rp, Morozova:2015bla}, the integrated flux can be written as in Eq.~\eqref{eq:lum_eq_ross} by formally defining the average
\begin{equation}
	\label{eq:R_Morozova}
	R_g = \frac{\int B_\nu R_\nu \,d\nu}{\int B_\nu \,d\nu} = \frac{\left| \nabla T^4 \right|}{\kappa_R \rho T^4} = \frac{4\pi r^2}{\kappa_R T^4} \left| \frac{\partial T^4}{\partial m} \right| \,,
\end{equation}
where we used the formal definition of the Rosseland mean opacity, $\kappa_R^{-1} = \int_{0}^{\infty} \kappa_\nu^{-1} \partial_T B_\nu \, d\nu \, / \! \int_{0}^{\infty} \partial_T B_\nu \, d\nu$, and the normalization $4 \pi \int_{0}^{\infty} B_\nu\, d\nu = ac\, T^4$.
The rightmost expression in Eq.~\eqref{eq:R_Morozova} only holds in spherical symmetry.
Inserting Eq.~\eqref{eq:R_Morozova} in place of $R_\nu$ into Eq.~\eqref{eq:flux_limiter} and then plugging the resulting gray $\lambda$ into Eq.~\eqref{eq:flux_eq} in place of $\lambda_\nu$ gives Eq.~\eqref{eq:lum_eq_ross}.
For our frequency-dependent radiation-hydrodynamic models, we get Eq.~\eqref{eq:R_factor} from Eq.~\eqref{eq:R_vector}, and
\begin{equation}
	L = - (4\pi r^2)^2 \left(\int_{0}^{\infty} \frac{4\pi \lambda_\nu}{3 \kappa_\nu} \frac{\partial B_\nu}{\partial T} d\nu \right) \frac{\partial T}{\partial m} \,
\end{equation}
from Eq.~\eqref{eq:flux_eq} and \eqref{eq:lum_eq}.

\section{Network initialization}
\label{app:res_NSEinit}

\begin{figure*}
	\centering
	\includegraphics[width=\textwidth]{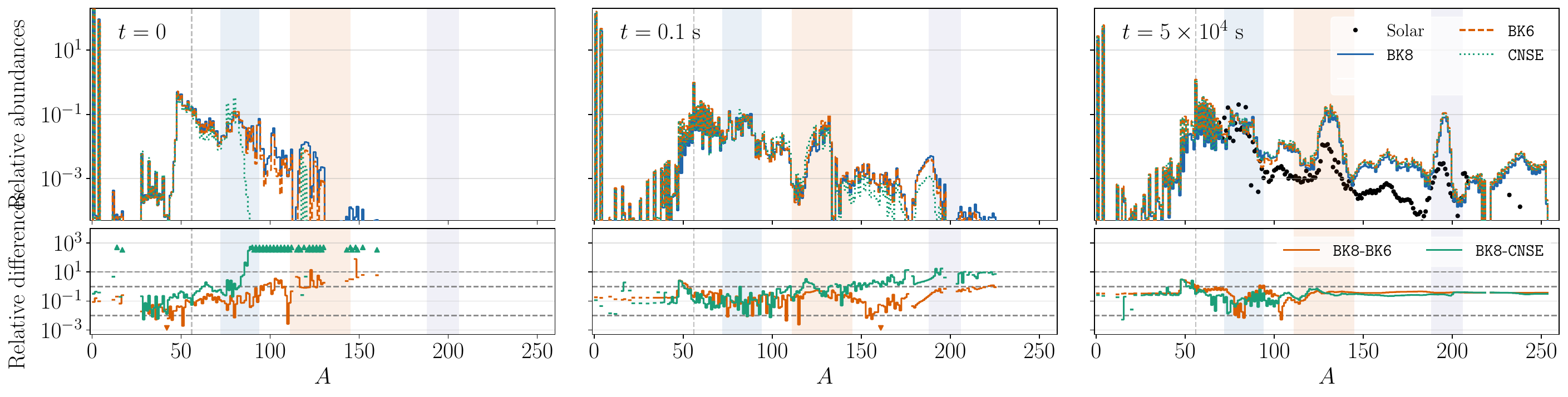}
	\caption{\textit{Top panels}: Mass-weighted composition of the \BLh\ ejecta profile after initialization (\textit{left column}), at $t=0.1$~s (pre neutron freeze-out, \textit{central column}), and $t=5\times10^4$~s (\textit{right column}).
		Blue, orange, and green lines show results respectively from the backtrack ($T_\text{NSE} = 8$~GK and $T_\text{NSE} = 6$~GK) and the cold-NSE initialization procedures.
		The black dots represent the Solar r-abundances \citep{Prantzos:2020a}.
		The abundances are scaled to match a unitary cumulative abundance for the elements of the first $r$-process peak, with $72 \leq A \leq 94$.
		The vertical dashed lines highlight the position of the elements $A=56$.
		The blue, orange, and purple shaded regions remark the position of the first, second and third $r$-process peaks, respectively.
		\textit{Bottom panel}: relative differences against the \texttt{BK8} run. The upward [downward] triangles indicate discrepancies greater than two orders of magnitude [smaller than $5\times10^{-4}$]. The horizontal dashed lines mark the factors $0.01$, 1, and 10 discrepancies.}
	\label{fig:yields_NSE}
\end{figure*}
\begin{figure*}
	\centering
		\includegraphics[width=\textwidth]{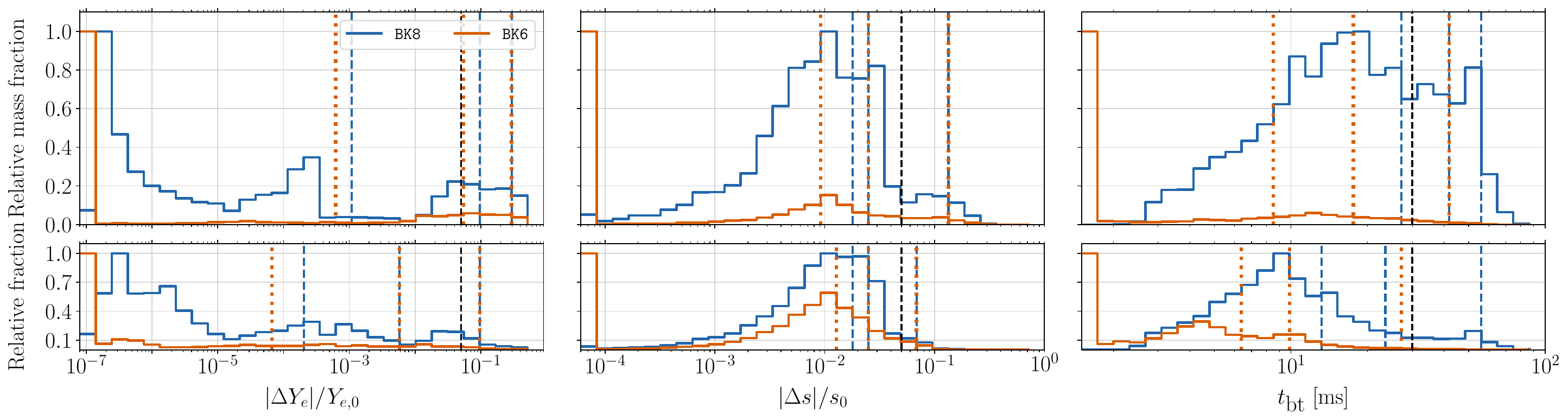}
	\caption{Histograms of the discrepancies between the extraction and post-backtracking electron fractions (first column) and entropies (second column), and pre-evolution time $t_\textrm{bt}$ (third column) for the \BLh\ ejecta profile. The colors represent different prescriptions for $T_\text{NSE}$. The histograms are normalized to the value of the highest bin.
		The vertical, colored lines mark the cumulative $0.75$, $0.9$ and $0.99$ (mass) fraction of the ejecta. The black dashed line correspond to a $5\%$ discrepancy in the first two columns, or a $30$~ms pre-evolution time.}
	\label{fig:bt_hist}
\end{figure*}

To investigate the systematic uncertainties on the nucleosynthesis arising from the NSE-based initialization procedures, we perform the same simulation three times, initializing the nuclear composition either with the backtracking method at $T_\text{NSE} = 6,8$~GK (\texttt{BK6} and \texttt{GK} simulations), or directly at the temperature recorded at extraction (\texttt{CNSE}).

In Fig.~\ref{fig:yields_NSE}, we show the composition for the \BLh\ ejecta profile at the beginning of the simulation, during the first phase of $r$-process nucleosynthesis, and once the production of the heavy nuclei is complete.
The cold-NSE initialization misses the early reactions after NSE drop-out (physically happening for many fluid elements during NR simulation times, $t < 0$).
Moreover, the low temperatures at extraction cause the Boltzmann term in the NSE equation to be overestimated (see Eq.~\eqref{eq:NSEcomp}).
Together, these two effects suppress or inflate the initial abundances of the isotopes around $A\sim65$ and $A\sim80$, respectively, and fail to capture the very early production of elements above the first peak.
The \texttt{BK6} and \texttt{BK8 }initializations are considerably more similar to each other, with differences from the lower temperature qualitatively similar to those discussed for the cold-NSE approach.
After initialization and during the $r$-process nucleosynthesis ($1~\mathrm{ms} \lesssim t \lesssim 1$~s), the rates of the dominant nuclear reactions are set by the local composition.
These rates evolve in a way that counteracts the initial errors, progressively reducing the systematic uncertainty over time.
As a result, the system gradually loses memory of the initialization method, with relative discrepancies in the late-time yields falling below unity for most mass numbers.
This effect is reflected in the final lanthanides mass fractions reported in Table~\ref{tab:knecnn_runs}, which agree within a few percent across the \texttt{GK}, \texttt{BK6}, and \texttt{CNSE} simulations.
As a notable exception, a group of isotopes slightly lighter than the iron group is overproduced by the more approximated initialization methods.
Single isotopes can experience more significant discrepancies. The cold-NSE method tends to leave less free neutrons after freeze-out, due to a stronger initial presence of seed nuclei. In our case, this introduces only minor corrections to the light curves, and only for $t \lesssim 0.5$~days.
					
The backtracking at the higher temperature $T_\text{NSE} = 8$~GK has in general to be preferred over the alternatives, as it approximately takes into account the otherwise overlooked early out-of-NSE processes.
However, backtracking initialization methods assume that the pre-evolution is short enough ($t \ll 1$~s) to have an approximately negligible effect on the electron fraction and entropy. We check these hypotheses in Fig.~\ref{fig:bt_hist}.
For the high initialization temperature, around $90\%$ of the ejected mass shows relative discrepancies respectively on the electron fraction and entropy below $10\%$ or $3\%$ after the pre-evolution. For around $75\%$ of the mass, this evolution lasts less than $30$~ms.
The backtracking assumptions are better satisfied by the low-temperature case.
For most of the ejecta, the backtracking is safe and closes before the ignition of the core reactions of the $r$-process nucleosynthesis.
Few percent of the fluid elements register pre-evolution times of $30$~ms or more, and a change in the $Y_e$ higher than $10\%$.
Such a discrepancy can be crucial for accessing the strong $r$-process nucleosynthesis for $Y_e \sim 0.22$ and to distinguish between neutron and proton rich material at $Y_e \sim 0.5$.
Long pre-evolution times also mean missing the NN-hydrodynamics coupling in the first phases of the $r$-process nucleosynthesis.
Figure~\ref{fig:yields_NSE} provides an estimate of the systematic uncertainties propagated from our initialization methods.
In the future, we plan to reduce these uncertainties by initializing the fluid element compositions using tracer particles extracted from the original NR data and preprocessed up to the initial time of the hydrodynamic simulation.

\end{appendix}

\end{document}